\documentclass[11pt]{article}
\usepackage[margin=1in]{geometry}
\usepackage{fullpage}
\usepackage[numbers,sort]{natbib}
\usepackage[utf8]{inputenc}
\usepackage{amsmath, amsthm, amssymb}
\usepackage{thmtools}
\usepackage{thm-restate}
\usepackage{amsfonts, bm, bbold, dsfont}
\usepackage{enumitem}
\usepackage[normalem]{ulem}
\usepackage{xspace}
\usepackage[usenames, dvipsnames]{xcolor}
\usepackage[colorlinks=true, citecolor=ForestGreen, linkcolor=ForestGreen, urlcolor=NavyBlue]{hyperref}
\usepackage{booktabs}
\usepackage{multirow}
\usepackage{relsize}
\usepackage{float}
\usepackage[nameinlink, capitalize]{cleveref}

\usepackage{tabularx}
\newcolumntype{Y}{>{\raggedright\arraybackslash}X}

\usepackage{comment}
\usepackage{subcaption}
\usepackage{multirow}
\usepackage{bbold}
\usepackage{pgf, tikz, pgfplots}
\usepackage{nicefrac}
\usepackage{bbm} 
\usepackage{placeins}
\usepackage[ruled,noend]{algorithm2e}
\crefname{algocf}{algorithm}{algorithms}
\Crefname{algocf}{Algorithm}{Algorithms}

\usetikzlibrary{matrix}
\usetikzlibrary{calc, patterns, intersections, pgfplots.fillbetween}
\pgfplotsset{compat=1.18}
\usetikzlibrary{arrows.meta,fillbetween} 

\usepackage[]{color-edits}

\newtheorem{theorem}{Theorem}[section]
\newtheorem{proposition}[theorem]{Proposition}
\newtheorem{lemma}[theorem]{Lemma}

\newtheorem{corollary}{Corollary}[theorem]

\theoremstyle{definition}

\allowdisplaybreaks
\SetAlFnt{\small}
\SetAlCapFnt{\small}
\SetAlCapNameFnt{\small}
\SetAlCapHSkip{0pt}
\IncMargin{-\parindent}

\usepackage{mathtools} 

\title{Online Maximin-Share Allocation with MMS Advice\thanks{The work of A. Berlin and A. Eden was supported by the Israel Science Foundation (grant No. 533/23). Alon Eden is the Incumbent of the Harry $\&$ Abe Sherman Senior Lectureship at the School of Computer Science and Engineering at the Hebrew University.}}

\author{Asnat Berlin \thanks{Hebrew University (\texttt{asnat.berlin@mail.huji.ac.il, alon.eden@mail.huji.ac.il}).} \and
Alon Eden \footnotemark[2] \and
Yaara Feldman \thanks{Weizmann Institute of Science (\texttt{yaara.feldman@weizmann.ac.il}).}}

\date{}

\begin{document}
\maketitle
    \setcounter{page}{1}
\thispagestyle{empty}

\pagestyle{empty}
\begin{abstract}
We study online fair division of indivisible goods with additive
valuations. Goods arrive one at a time or in batches, and each batch
must be allocated irrevocably before the next one is revealed.
The algorithm is given the agents' maximin share (MMS) values in
advance, or estimates thereof. This advice is motivated by settings
such as repeated food donations, where past allocations can inform
estimates of future fairness benchmarks.

For heterogeneous valuations, we determine the asymptotically optimal
MMS approximation, including the leading constant. We give an algorithm that obtains an $(1-o(1))/H_n$-approximation for item-by-item arrivals and $(1-o(1))/H_{\min\{k,n\}}$ for $k$ batches as $\min\{k,n\}\to\infty$, where $n$ is the number of agents and $H_\ell\approx \ln \ell$ is the $\ell$th harmonic number. We provide asymptotically tight impossibilities by showing that no algorithm with exact MMS advice can approximate the MMS better than  $1/H_n$ and $(1+o(1))/H_{\min\{k,n\}}$-approximation for the item-by-item and batched arrival models, respectively. Both asymptotic upper bounds hold even when the algorithm knows every agent's exact MMS and total value
in advance. We sharpen the
upper bound to $12/19$ for two agents. For two batches, we give a
$4/7$-approximation and rule out a $2/3$-approximation in general.

For identical valuations, we show that the optimal approximation
for two batches is exactly $2/3$. For $k\geq 3$, we give a
$(k+1)/(2k+1)$-approximation and establish complementary
impossibility bounds for three, four, and six batches.
Our impossibility results hold even with exact MMS advice, and
our algorithmic guarantees degrade gracefully when the advice
is approximate.
\end{abstract}
\newpage
    \setcounter{tocdepth}{2} 
    \tableofcontents

    \newpage

    \pagestyle{plain}
    \pagenumbering{arabic}
\section{Introduction}

Maximin share (MMS), introduced by~\citet{Budish11, hill1987partitioning}, is the ubiquitous share-based fairness notion for allocating indivisible goods among agents.  The MMS
of an agent is the largest value that the agent can guarantee by partitioning
all goods into $n$ bundles and receiving the least valuable bundle.  Exact
MMS allocations need not exist for more than two agents
\citep{KurokawaProcacciaWang18}. For this reason, approximate MMS allocations have been extensively studied~\cite{KurokawaProcacciaWang18,HeidariEtAl26,HuangZhou25}. For $\alpha<1$, an $\alpha$-approximate allocation is an allocation where every agent gets at least an $\alpha$-fraction of their MMS value. We focus on additive valuations. For such valuations, approximate MMS allocations are known to exist with a constant $\alpha$, with the state-of-the-art approximation of $\frac{7}{9}$~\cite{HuangZhou25}. On the negative side, \citet{FeigeSapirTauber22} showed a three-player instance in which every allocation gives at least one agent no more than a $39/40$-fraction of their MMS value. \citet{EzraGarbuz26} give an offline upper bound of $20/21$ for
four agents and an upper bound of $1-\Omega(1/(\log n)^2)$ for
sufficiently large $n$.

In this paper, we study settings, such as food donations~\cite{Aleksandrov15,BenadeEtAl24} and time slots on shared computing clusters~\cite{GaoPeysakhovichKroer21,MahajanEtAl20}, where goods become available piecemeal over time, either item-by-item or in batches. Thus, currently available goods must be allocated before future arrivals are known due to storage constraints, expiration dates, or scheduling commitments. 
In online settings, the benchmark depends on the complete instance, while the allocation
must be constructed through irrevocable decisions. Indeed, \citet{ZhouBaiWu23} show that in this setting, no finite approximation is achievable, even for three players and at most six items. Even when goods arrive in only two batches, the optimal deterministic
MMS approximation ratio without prior information is exactly $1/n$;
the impossibility holds even for identical valuations
(Appendix~\ref{app:two-batches-no-advice}).

This tension raises an informational question: how much of the difficulty of
online MMS allocation is caused by not knowing the value of the benchmark
itself? Indeed, in many scenarios, past data can be used to infer the fairness criteria.  We study an advice model in which the MMS value of each agent, or an approximation thereof, is
known in advance to the algorithm. The advice reveals neither values of future goods
nor a partition witnessing the MMS values. Thus, the algorithm knows the
fairness target that it should approximate, but must still decide online how to
assemble bundles that meet that target. This assumption was made for the online machine covering problem, which is equivalent to MMS allocation with identical valuations~\cite{online-covering97,semi-online-covering}. In this paper, we study the power of online algorithms with MMS advice in online settings where the items arrive either item-by-item or in batches, both for heterogeneous and identical valuations.

\paragraph{Our contributions.}

\textit{Heterogeneous valuations (\Cref{sec:heterogeneous}).}
Let $\tilde k=\min\{k,n\}\to\infty$, where $n$ is the number of agents and $k$ is the number of batches.  We give an $(1-o(1))/H_n$-approximation under item-by-item arrivals
and an $(1-o(1))/H_{\tilde k}$-approximation with $k\ge2$ batches
(\Cref{thm:unified-heterogeneous}). Our algorithm combines priority-based
bundle completion with weighted allocation of the remaining goods. The
weights exploit the fact that at most one residual allocation is made per
batch, yielding a guarantee that improves with fewer batches.

We show that the logarithmic loss is unavoidable, matching these
guarantees asymptotically, including the leading constant.
No algorithm can guarantee more than $1/H_n$ under item-by-item
arrivals, or $(1+o(1))/H_{\tilde k}$ with $k$ batches as
$\tilde k\to\infty$
(\Cref{thm:item-by-item-bound,thm:batched-bound}).
Both bounds hold even when every agent's MMS and total value are
known in advance. Our constructions keep allocated values
small while maintaining a separate MMS partition for each agent.
A harmonic condition on accumulated tentative values allows these
partitions to be completed while ensuring that some agent receives
no additional value.

For two agents, we strengthen the item-by-item upper bound
to $12/19$ (\Cref{sec:impossibility-2}).
For two batches, we obtain a $4/7$-approximation, while no
algorithm can exceed $2/3-1/2322429$
(\Cref{thm:two-batch-possibility,thm:two-batch-impossibility}).

\textit{Identical valuations (\Cref{sec:identical}).}
For $k\ge3$ batches, we obtain $(k+1)/(2k+1)$
(\Cref{thm:identical-k-batches}). 
Our algorithm keeps at most one unfinished bundle and bounds the weight of
completed bundles relative to an MMS partition.

We prove impossibility bounds of $12/19$ for three batches, 
$3/5$ for four,  
and $4/7$ for six 
in \Cref{sec:impossible-c-batches}. The constructions adapt to
previous allocations, leaving too few goods to complete all remaining
agents. For four and six batches, we adapt machine-covering constructions~\cite{online-covering97}.

For two batches, the optimal guarantee is $2/3$
(\Cref{sec:results-identical-2batches}). We maximize the number of completed
bundles and concentrate the remaining goods on one agent. A two-branch
machine-covering construction gives the matching impossibility.

Positive results use only MMS lower bounds: a guarantee of $\alpha$ becomes $\alpha\eta^2$ when the prediction is accurate up to an $\eta\in (0,1]$ factor (see \Cref{prop:approx-advice}).

\vspace{2mm}

Our impossibility result rules out a constant approximation for heterogeneous valuations under item-by-item arrivals, even when both the MMS values and the total values are known in advance.

\begin{table*}[t]
\centering
\caption{Summary of MMS approximation guarantees. An achievable bound is a
lower bound on the approximation factor; an impossibility bound is an upper
bound on what any online algorithm can guarantee.}
\label{tab:main-results}
\setlength{\tabcolsep}{4pt} 
\renewcommand{\arraystretch}{1.18}
\begin{tabularx}{\textwidth}{@{}p{0.18\textwidth}p{0.15\textwidth}YY@{}}
\toprule
Valuations & Arrival model & Achievable guarantee & Impossibility bound \\
\midrule

Heterogeneous
& Item by item
& $2-\sqrt{2}$ for $n=2$; $\alpha_n\sim (1-o(1))/H_n$
  [\Cref{thm:unified-heterogeneous}]
& $12/19$ for $n=2$ [\Cref{thm:supp-two}];
  $1/H_n$ for general $n$
  [\Cref{thm:item-by-item-bound}] \\

Heterogeneous
& $k$ batches
& $4/7$ for $k=2$ [\Cref{thm:two-batch-possibility}]; $\beta=\min\{1/2,\alpha_{\tilde k}\}$ for $n\ge3$
  [\Cref{thm:unified-heterogeneous}]
& $<2/3$ for $k=2,\ n\ge 6$ [\Cref{thm:two-batch-impossibility}]; $(1+o(1))/H_{\tilde k}$ for $\tilde k\to\infty$ [\Cref{thm:batched-bound}]\\

\midrule

Identical
& Item by item
& $\frac{n}{2n-1}$ for $n\in\{2,3,4\}$
  \citep{online-covering97};
  $6/11$ for general $n$
  \citep{semi-online-covering}
& $\frac{n}{2n-1}$ for $n\in\{2,3,4\}$ \citep{online-covering97};
  $24/43$ for general $n$
  \citep{semi-online-covering} \\

Identical
& $2$ batches
& $2/3$
  [\Cref{sec:results-identical-2batches}]
& $2/3$
  [\Cref{sec:results-identical-2batches}] \\

Identical
& $3$ batches
& $4/7$
  [\Cref{cor:combined-identical-k-batches}]
& $12/19$ for $n\ge38$
  [\Cref{lem:three-batches-twelve-nineteenths}] \\

Identical
& $4$--$5$ batches
& $5/9$ for $k=4$; $6/11$ for $k=5$
  [\Cref{cor:combined-identical-k-batches}]
& At most $3/5$ for $n\geq3$
  [\Cref{lem:imp-4batches}] \\

Identical
& $k\geq6$ batches
& $6/11$
  \citep{semi-online-covering}
& At most $4/7$ for $n\geq4$
  [\Cref{lem:imp-6batches}] \\

\bottomrule
\end{tabularx}

\vspace{3pt}
\parbox{\textwidth}{\footnotesize
Here $\tilde k=\min\{k,n\}$.
Let $c_t$ is the positive solution of
$\sum_{r=1}^{t}(1+c_t r)^{-1}=1$, and
$\alpha_t=(1+c_t)^{-1}$. Bounds for four and six batches extend to larger
numbers of batches by padding with zero-valued batches.}
\end{table*}


\paragraph{Relations to other advice models.}
Our advice consists of a single value per agent, but the choice of this
value matters. Knowing each agent's total value permits a tight
$1/2$-MMS guarantee for two agents and no positive MMS guarantee for
$n\ge3$ \citep{ZhouBaiWu23,ChenTan26}, although it suffices for PROP1
\citep{additiniol-information25}. Maximum-item-value advice gives
$1/n$-PROP1 \citep{inapprox-mms26}, improved to $19/30$ by
\citet{NeohTeh26} and $1/2$ by \citet{DuSun26}.
These are guarantees of proportionality \emph{up to one good}: an agent
may add a good outside their bundle when evaluating the guarantee.
An $\alpha$-approximation to proportionality implies $\alpha$-MMS, since
$\mu_i\le v_i(M)/n$, but an $\alpha$-PROP1 guarantee does not.
Indeed, even exact maximum-item-value advice admits no positive MMS
approximation \citep{inapprox-mms26}. In contrast, MMS advice yields
$\Omega(1/\ln n)$-MMS for every $n$ (\Cref{thm:unified-heterogeneous}).
Neither totals nor maximum values determine MMS, and MMS does not determine
either of them. For identical valuations, MMS advice coincides with
knowing the optimal minimum load in machine covering
\citep{online-covering97,semi-online-covering}. This permits a $6/11$
guarantee, whereas knowing only the total load or only the largest item
gives a tight $1/(n-1)$ guarantee for $n\ge3$ \citep{TanWu07}.

Frequency advice provides substantially more information: for each agent,
it specifies the entire multiset of future values, without their arrival
order \citep{additiniol-information25}. This determines the agent's MMS,
total and maximum values. Our advice retains only the MMS
and need not describe the future goods or an MMS partition.
\citet{additiniol-information25} show that exact frequency advice
transfers any feasible offline share guarantee to the online setting.
Combined with \citet{HuangZhou25}, this yields a $7/9$-MMS guarantee
for every $n$. Recent independent results by \citet{FeigeGrinberg26}
and \citet{LinEtAl26} show that the optimal offline MMS approximation
for goods tends to $1$ as $n$ grows. The same therefore holds online
with exact frequency advice. Under item-by-item arrivals, our upper
bound of $1/H_n$ tends to $0$ even when the algorithm knows every
agent's exact MMS and total value in advance
(\Cref{thm:item-by-item-bound}).

Both models also admit guarantees that
degrade with prediction error \citep{additiniol-information25}. 

\paragraph{Further Related Work.}

\textit{Online fair division with additional information.}
Earlier work studies how advice improves online matching and welfare
\citep{AleksandrovWalsh17}. Other work uses predictions to obtain EFX
\citep{MelissourgosProtopapas25} or envy guarantees under budget constraints
\citep{CohenEtAl26Budget}. Some models give the algorithm limited
information about upcoming goods \citep{AmanatidisEtAl25}, or allow it to
use a buffer to postpone allocation 
\citep{AmanatidisEtAl26Buffer}. Our algorithms allocate the current
batch before seeing the next one.

\textit{Information and predictions in other fair division settings.}
For divisible goods, several works study fairness and welfare when agents'
total values are known \citep{GkatzelisPsomasTan21,BarmanKhanMaiti22,HuangEtAl25Welfare}.
Predictions of these totals are also used to approximate Nash social
welfare \citep{BanerjeeEtAl22} and proportional fairness for public goods
\citep{BanerjeeEtAl23}. Other work uses predicted agent weights
\citep{CohenPanigrahi23} or workload forecasts for allocating content
moderation tasks \citep{Allouah23}. Predictions and prior information also
play a role in truthful fair allocation
\citep{CohenEtAl24,GkatzelisEtAl24Information}.
In a different online model, agents arrive over time, and the algorithm
knows their possible valuation functions \citep{KulkarniMehtaShahkar25}.
In this agent-arrival model, \citet{ZhouKnownTypes26} show that
the optimal deterministic MMS guarantee is $\Theta(1/\sqrt{t})$
when there are $t$ known valuation types.
Temporal fair division asks for fairness throughout the sequence of
allocations \citep{ElkindEtAl25,CooksonEbadianShah25}.
Further work studies fair division with stochastic valuations
\citep{ZengPsomas20}, rankings of goods \citep{BenadeHalpernPsomas25},
inaccurate valuations \citep{Murray2020,DangEtAl26}, or values learned
from feedback during allocation \citep{YamadaEtAl24,ProcacciaSchifferZhang24}.

\textit{Online fair division without predictions.}
\citet{Aleksandrov15} study online allocation motivated by food banks;
\citet{AleksandrovWalsh20} survey early work.
\citet{BenadeEtAl18} bound envy as goods arrive, allowing goods to arrive
in batches, and \citet{HalpernEtAl25Envy} give further bounds for online
envy minimization.

Other work studies the tradeoff between fairness and
welfare \citep{BenadeEtAl24}, or market-based allocation when goods arrive
from an unknown distribution \citep{GaoPeysakhovichKroer21}.
Positive results also follow from restricting valuations
\citep{AmanatidisEtAl25,WangWei26} or allowing guarantees to depend on the
number of goods \citep{DuSun26}. \citet{CohenTehWooldridge26} study the
case where the arrival sequence is fixed before the algorithm runs.
Other models allow changing earlier allocations \citep{HeEtAl19} or
paying subsidies to eliminate envy \citep{KulkarniEtAl26Subsidy}.
For arriving agents, \citet{ChawlaEtAl26} obtain a constant MMS
approximation for all agents with constant probability, without prior
valuation information, when the arrival sequence is fixed in advance.
Offline MMS approximation is studied by
\citet{KurokawaProcacciaWang18,HeidariEtAl26,HuangZhou25}.


\section{Preliminaries}

Let $N=[n]$, where $n\geq2$, be the set of agents. The number $n$ of agents
is known to the algorithm in advance. We consider two arrival models.
In the \emph{item-by-item model}, goods arrive one at a time, and each good
must be allocated before the next is revealed. In the \emph{batched model},
goods arrive in $k$ batches, where $k$ is known in advance. The ground set
$M$, the number of goods in each batch, and all future values are initially
unknown.

In the batched model, the goods arrive in $k$ disjoint batches
$B_1,\ldots,B_k$, whose union is $M$. When batch $B_t$ arrives, the
algorithm observes the complete value vector $\bigl(v_i(g)\bigr)_{i\in N}$ of every $g\in B_t$. It must allocate every good in $B_t$ before
$B_{t+1}$ is revealed. The item-by-item model is the special case in which
each batch contains a single good, without advance knowledge of the number
of goods. Allocations are irrevocable in both models.

Each agent $i$ has a nonnegative additive valuation, that is $v_i(S)=\sum_{g\in S}v_i(g)$
for every $S\subseteq M$. After batch $t$, the allocation is a tuple
$(A_1^t,\ldots,A_n^t)$ that partitions
$\bigcup_{s\leq t}B_s$ and satisfies $A_i^{t-1}\subseteq A_i^t$ for every $i\in N$. We write $A_i=A_i^k$ for the final allocation of agent $i$.

\paragraph{Maximin Share Fairness.}
For each agent $i$, define the maximin share (MMS) with respect to $n$ agents and the full item set $M$ as
\[
\mu_i(n,M) 
:= 
\max_{\Pi \in \mathcal{P}_n(M)} 
\min_{S \in \Pi} v_i(S),
\]
where $\mathcal{P}_n(M)$ denotes the set of all partitions of $M$ into $n$ bundles.

We write $\mu_i=\mu_i(n,M)$ when $n$ and $M$ are clear from context.

\paragraph{MMS advice.}
Before any goods arrive, the algorithm is given the exact values
$\mu_i(n,M)$, but not a partition attaining them. We omit agents with zero MMS, whose guarantee is automatic, and
normalize the remaining agents' original MMS values to $1$,
which is still a valid lower bound on their MMS after the removal. We also consider \textit{Approximate MMS advice} $\widehat\mu_i$ with accuracy parameter
$\eta\in(0,1]$ satisfying 
$$\eta\mu_i\le\widehat\mu_i\le\mu_i/\eta.$$

For identical valuations, if several $\eta$-accurate estimates of the common MMS are given, we use $\eta\max_{i\in N}\widehat\mu_i$ as our
common lower bound.

\paragraph{Approximate MMS}
An algorithm is said to achieve an $\alpha$-approximation to MMS if, for every agent $i$,
$v_i(A_i) \ge \alpha \cdot \mu_i(n,M).$

We next claim that running an approximation algorithm with an approximate MMS advice retains approximation guarantees. We say an $\alpha$-approximation algorithm is \textit{downward-compatible} if the advice it receives satisfies $\underline\mu_i\le \mu_i$ for every agent $i$, then every agent $i$'s allocated bundle $A_i$ guarantees that $v_i(A_i)\ge \alpha \underline\mu_i$. We note that all of the algorithms throughout the paper are trivially \textit{downward compatible}. We have the following.

\begin{proposition} \label{prop:approx-advice}
    Let $A$ be a downward-compatible $\alpha$-approximation algorithm to the MMS when the advice is accurate, and let $\hat \mu_i$ be an $\eta$-accurate estimate. Then using $A$ with $\underline\mu_i=\eta\widehat\mu_i$ as the advice guarantees $\alpha\eta^2$-approximation to the MMS. 

\end{proposition}
\begin{proof}
Suppose  that
$\eta\mu_i\le\widehat\mu_i\le\mu_i/\eta$.
Then $\eta^2\mu_i
    \leq
    \eta\widehat\mu_i
    \leq
    \mu_i.$
Thus $\underline\mu_i=\eta\widehat\mu_i\le \mu_i$. Since the algorithm is downward-compatible,
applying the original guarantee gives
\[
    v_i(A_i)
    \geq
    \alpha \underline\mu_i
    =
    \alpha\eta\widehat\mu_i
    \geq
    \alpha\eta^2\mu_i.
\]

If instead
$\eta\mu_i\leq\widehat\mu_i\leq\mu_i$, then
$L_i=\widehat\mu_i$ is already a valid lower bound, and therefore
\[
    v_i(A_i)
    \geq
    \alpha\widehat\mu_i
    \geq
    \alpha\eta\mu_i.
\]
\end{proof}



\section{Heterogeneous Valuations} \label{sec:heterogeneous}

\subsection{$(1-o(1))/H_{\min\{n,k\}}$-approximation}
\label{sec:logarithmic-guarantee}

In this section, we give an algorithm whose approximation guarantee depends
logarithmically on the minimum between the number of batches and the number of agents. The algorithm builds bundles towards the target approximate MMS value. Whenever the current batch contains a set that brings some agent to
the target, the algorithm allocates such a set, giving priority to agents
with smaller indices. The main difficulty is deciding who should receive
the remaining items when they do not suffice to complete any agent.

To allocate these remaining items, we assign each agent a weight and choose
an agent with maximum weighted value for the set. Agents with larger
indices receive larger weights, compensating for their lower priority when
bundles are completed. The choice of weights uses the fact that at most one
such residual allocation is made per batch. Thus, at most $\min\{k,n\}$
agents receive residual items, even if the number of goods is unbounded.
We choose the weights so that the weights of any such set of agents sum to
at most $1$. This will bound the value of all residual allocations from
the perspective of an agent who does not reach the target.

For every integer $t\ge2$, define
\[
    F_t(c)=\sum_{h=1}^t\frac{1}{1+ch}.
\]
Let $c_t$ be the unique positive solution of $F_t(c_t)=1$, and set
$\alpha_t=1/(1+c_t)$. The function $F_t$ is strictly decreasing in $c$
and increasing in $t$; hence $c_t$ is increasing and $\alpha_t$ is
decreasing in $t$. The following technical lemma regarding the properties of the $\alpha_t$s will be used in the approximation analysis.

\begin{lemma}\label{lem:weighted-harmonic}
For every $t\ge2$,
\[
    \frac{1}{1+H_{t-1}}<\alpha_t<\frac{1}{H_t},
\]
where $H_t=\sum_{h=1}^t1/h$ is the $t$-th harmonic number.
Moreover, $\alpha_2=2-\sqrt2$.
\end{lemma}
\begin{proof}
For $h\le t$,
\[
    1+H_{t-1}h=H_th+1-h/t\ge H_th,
\]
with strict inequality when $h<t$. Thus,
\begin{align*}
    F_t(H_{t-1})
    &=\sum_{h=1}^t\frac{1}{1+H_{t-1}h}<\sum_{h=1}^t\frac{1}{H_th}=1.
\end{align*}
Since $F_t$ is decreasing, $c_t<H_{t-1}$, proving the lower bound on
$\alpha_t$. For the upper bound, $1+c_th\le(1+c_t)h$ implies
\[
    \frac{1}{1+c_th}\ge\frac{1}{(1+c_t)h}
    =\frac{\alpha_t}{h},
\]
with strict inequality when $h>1$. Since $t\ge2$, summing gives
\begin{align*}
    1=F_t(c_t)
    &=\sum_{h=1}^t\frac{1}{1+c_th}
    >\alpha_t\sum_{h=1}^t\frac{1}{h}
    =\alpha_t H_t.
\end{align*}

Solving $F_2(c_2)=1$ gives $\alpha_2=2-\sqrt2$.
\end{proof}

Let $\tilde k=\min\{k,n\}$ and
$\beta=\min\{1/2,\alpha_{\tilde k}\}$.
For batched arrivals and $n\ge3$, we use target $\beta$ and weights
\[
    w_i=\frac{1}{1+c_{\tilde k}(n-i+1)}.
\]
The equation defining $c_{\tilde k}$ says precisely that the
$\tilde k$ largest weights sum to $1$. Under item-by-item arrivals, 
we use target $\alpha_n$ and the same weights with $c_n$ in place of
$c_{\tilde k}$. In this case, all $n$ weights sum to $1$.
For $n=2$, we use this item-by-item implementation within each batch,
processing its goods in an arbitrary fixed order.

We present the allocation rule for a target $\tau>0$ and positive weights
$w_1,\ldots,w_n$. An agent is \emph{active} while the value of their bundle
is below $\tau$. A set \emph{completes} an active agent if adding it to
their bundle brings its value to at least $\tau$.
The algorithm first allocates goods that individually complete an agent.
It then scans the remaining goods and allocates the shortest prefix that
completes an agent. If no such prefix exists, it assigns the remaining
goods according to their weighted values. The formal description appears
in \Cref{alg:weighted-allocation}.

\begin{algorithm}[t]
\caption{Weighted allocation with target $\tau$ and weights $w_1,\ldots,w_n$}
\label{alg:weighted-allocation}
Initially, every agent has an empty bundle and is active. For each arriving
batch, let $R$ be its unallocated goods and apply the following rules while
$R\neq\emptyset$ and an active agent remains.
\begin{enumerate}
    \item While some $g\in R$ completes an active agent, give $g$ to the
    smallest-index active agent it completes. Remove $g$ from $R$, declare
    its recipient inactive.

    \item Otherwise, scan $R$ in a fixed order. While a prefix completes an
    active agent, give the shortest such prefix to the smallest-index
    active agent it completes. Remove the prefix from $R$, declare its
    recipient inactive.

    \item Otherwise, give all of $R$ to an active agent $j$ maximizing
    $w_jv_j(R)$ and proceed to the next batch.
\end{enumerate}
Inactive agents receive no further goods while an active agent remains.
\end{algorithm}

\begin{theorem}\label{thm:unified-heterogeneous}
With the parameters specified above,
\Cref{alg:weighted-allocation} guarantees an
$\alpha_n$-approximation under item-by-item arrivals and a
$\beta$-approximation under $k\ge2$ batched arrivals for $n\ge3$.
For $n=2$, its item-by-item implementation guarantees
$\alpha_2=2-\sqrt2$ in both arrival models.
In particular, the guarantees are $(1-o(1))/H_n$ as $n\to\infty$
and $(1-o(1))/H_{\min\{k,n\}}$ as $\min\{k,n\}\to\infty$,
respectively.
\end{theorem}

The approximation proof bounds separately the value of the residual allocations made by
Rule~3 and the sets that complete agents under Rules~1 and~2. The first
bound follows from the choice of weights. The second follows from the
priority order and the minimality of each completion set. Together, these
bounds imply that an agent who fails to reach the target cannot have MMS
equal to $1$, a contradiction.

For each agent $j$, let $P_j$ be the union of the sets assigned to $j$ by
Rule~3, and let $D=\{j:P_j\neq\emptyset\}$. If $j$ becomes inactive, let
$S_j$ be the set assigned to $j$ in the step that completes them.

\begin{lemma}
\label{lem:weighted-residual}
Suppose agent $i$ remains active throughout the execution. If
$\sum_{j\in D}w_j\le1$, then $\sum_{j=1}^n v_i(P_j)<\frac{\tau}{w_i}.$
\end{lemma}
\begin{proof}
Whenever Rule~3 assigns a set $R$ to agent $j$, agent $i$ is active, so
$w_iv_i(R)\le w_jv_j(R)$. Moreover, $j$ remains active after this
allocation, otherwise Rule~2 would have applied to $R$. Since $j$ receives
no further goods after becoming inactive, $v_j(P_j)<\tau$.
If $D=\emptyset$, the claim is immediate. Otherwise, summing over the
residual allocations gives
\[
    \sum_{j=1}^n v_i(P_j)
    \le \frac{1}{w_i}\sum_{j\in D}w_jv_j(P_j)
    < \frac{\tau}{w_i}\sum_{j\in D}w_j
    \le \frac{\tau}{w_i}.
\]
\end{proof}

The next lemma bounds the completion sets. Singleton sets assigned to
agents preceding $i$ may have arbitrarily large value to $i$. We account
for them directly using an MMS partition in the proof of the theorem.
\begin{lemma}
\label{lem:weighted-completion}
Suppose agent $i$ remains active throughout the execution. Every completion
set $S_j$ with $j>i$ satisfies $v_i(S_j)<\tau$. Every non-singleton
completion set $S_j$ with $j<i$ satisfies $v_i(S_j)<2\tau$.
\end{lemma}
\begin{proof}
If $j>i$ and $v_i(S_j)\ge\tau$, then $S_j$ also completes $i$. The
priority rule would therefore choose $i$ over $j$, a contradiction.

Now let $j<i$ and suppose $S_j$ is not a singleton. Let $g$ be the last
good in the prefix $S_j$. Rule~1 did not apply to $g$, so $v_i(g)<\tau$.
This remains true during Rule~2, since the bundles of agents who remain
active do not change during prefix completions. By minimality of the
chosen prefix, $S_j\setminus\{g\}$ does not complete $i$, and hence
$v_i(S_j\setminus\{g\})<\tau$. Adding the two bounds proves the claim.
\end{proof}

\begin{proof}[Proof of \Cref{thm:unified-heterogeneous}]
Suppose that some agent $i$ finishes with value below the target $\tau$.
Then $i$ remains active throughout the execution. In the batched
implementation, Rule~3 is used at most once per batch, so
$|D|\le\tilde k$. Since the $\tilde k$ largest weights sum to $1$,
\Cref{lem:weighted-residual} applies. It also applies to the item-by-item
implementation, where all $n$ weights sum to $1$.

Let $G$ be the set of singleton completion goods assigned to agents
preceding $i$, let $q=|G|$, and let $p$ be the number of non-singleton
completion sets assigned to agents preceding $i$. Write $r=n-i+1$.
Since $q+p\le i-1$, removing $G$ leaves at least
$n-q\ge r+p$ bundles of an MMS partition for $i$ intact. Therefore,
\[
    v_i(M\setminus G)\ge r+p.
\]
On the other hand, $M\setminus G$ consists of the residual allocations,
the $p$ non-singleton completion sets assigned before $i$, and at most
$r-1$ completion sets assigned after $i$. By
\Cref{lem:weighted-residual,lem:weighted-completion},
\begin{equation}\label{eq:weighted-value-bound}
    v_i(M\setminus G)
    <\frac{\tau}{w_i}+2\tau p+\tau(r-1).
\end{equation}

For batched arrivals with $n\ge3$, we have $\tau=\beta$ and
$w_i^{-1}=1+c_{\tilde k}r$. Thus, the right-hand side of
\eqref{eq:weighted-value-bound} equals
\[
    \beta(1+c_{\tilde k})r+2\beta p\le r+p,
\]
since $\beta\le\alpha_{\tilde k}$ and $\beta\le1/2$. This contradicts
the lower bound above. Moreover, $\alpha_{\tilde k}\ge\alpha_n$ and,
for $n\ge3$, $F_n(1)\ge F_3(1)=13/12>1$, so $\alpha_n<1/2$.
Hence $\beta\ge\alpha_n$. To obtain the stated bound, if
$\tilde k=2$, then $\beta=1/2=1/(1+H_1)$. If $\tilde k\ge3$, then
$F_{\tilde k}(1)\ge F_3(1)>1$, so $\beta=\alpha_{\tilde k}$.
By \Cref{lem:weighted-harmonic}, this is greater than
$1/(1+H_{\tilde k-1})=(1-o(1))/H_{\tilde k}$.

Under item-by-item arrivals, every completion set is a singleton, so
$p=0$. Substituting $\tau=\alpha_n$ and $w_i^{-1}=1+c_nr$ into
\eqref{eq:weighted-value-bound} gives
\[
    v_i(M\setminus G)<\alpha_n(1+c_n)r=r,
\]
a contradiction. By \Cref{lem:weighted-harmonic},
$\alpha_n\sim (1-o(1))/H_n$, proving the stated item-by-item guarantee.
For $n=2$, the same argument guarantees $\alpha_2=2-\sqrt2$, including
when we process each batch item by item. 
\end{proof}


\subsection{Impossibility Bounds}  \label{sec:impossible-n-agents-heterogeneous}

We next show that the algorithmic guarantees above are asymptotically
optimal, including the leading constant. For item-by-item arrivals,
we prove an upper bound of $1/H_n$. For $k$ batches, we prove an upper
bound of $(1+o(1))/H_{\tilde k}$ as $\tilde k=\min\{k,n\}\to\infty$.
Both bounds hold even when the algorithm knows every agent's exact
MMS and total value in advance.

The two constructions use the same completion argument. The adversary
builds a separate partition for each agent and arranges that some
agents need fewer completion goods than there are agents in their
group. This allows the adversary to complete the partitions while
leaving at least one of these agents with no additional value.

\subsubsection{Item-by-item}\label{sec:harmonic-impossibility}

\paragraph{Overview of the input sequence}

The construction has two phases.
In the first phase, the adversary gives small goods one at a time.
Whenever an agent receives a good, the adversary lowers their tentative
value for the next good. This keeps the value of their allocated bundle
small, even if they receive many goods.  To show that their MMS is at least~$1$, however, we can
use all the goods, including those allocated to other agents. The
adversary keeps a separate partition of these goods for each agent,
filling one bundle to value $1$ before starting the next. We switch
to the second phase as soon as there is a group of agents such that
each has fewer bundles left to fill than there are agents in the group.
The adversary then gives large goods to complete every agent's partition,
choosing their values so that the group has too few goods of positive
value to give one to each of its members.

More precisely, suppose that $n-h+1$ agents each have at least $h$
full bundles. Each needs at most $n-h$ more goods to complete their
partition. We can use the same $n-h$ goods for all of them, since a
good may have a different value for each agent. All other goods in
the second phase are worth zero to these agents. Thus one of them
finishes with only the value they received in the first phase, even
though they can partition all the goods into $n$ bundles of value $1$.

\noindent\textbf{First phase.}
Fix a sufficiently large integer $T$. We will show that the adversary
can switch phases after at most $T$ goods. Set
$L_T=\sum_{t=0}^{T-1}1/(n+t)$ and $D=H_n L_T-1$, so $D>1$ for
large enough $T$. After $t$ goods, let $x_i(t)=|A_i^t|$ be the number of items allocated to $i$ thus far. The tentative
value of the next good $g$ to agent $i$ is
\[
 q_i(g)=\frac{1}{D(1+x_i(t))}.
\]
For each agent $i$, the adversary keeps a partition of the goods
revealed in this phase into $n$ bundles, initially empty. Place $g$ in
the first bundle worth less than $1$ to $i$. If that bundle needs
$c$ more value to reach $1$, set $v_i(g)=\min\{q_i(g),c\}$.
If all bundles are full, set $v_i(g)=0$ and place $g$
in any bundle. The algorithm sees the resulting values before it
allocates $g$.

\noindent\textbf{Second phase.}
As soon as at least $n-h+1$ agents each have at least $h$ full bundles
for some $1\le h\le n$ (which is guaranteed to occur by \Cref{lem:switch}), introduce $n$ goods $g_1,\ldots,g_n$ one at a time.
For each $i\in N$, set $v_i(g_j)$ to the amount missing from their
$j$th bundle. Thus an empty bundle gets a good worth $1$, and placing
each $g_j$ in its corresponding bundle brings every bundle to $1$.

\paragraph{Proof of the item-by-item bound}\label{sec:item-by-item-proof}

For each agent $i$, let $R_i$ be the sum of their tentative values
for all goods revealed so far in the first phase.

\begin{lemma}\label{lem:switch}
After at most $T$ goods in the first phase, there is some $1\le h\le n$
such that at least $n-h+1$ agents each have at least $h$ full bundles.
\end{lemma}

\begin{proof}
If the condition holds before the $T$th good, we are done. Otherwise
the adversary gives all $T$ goods in the first phase. We first bound $1/R_i$ for a fixed agent $i$.
The tentative value of good $t+1$ to $i$ is $1/[D(1+x_i(t))]$.
Summing over the $T$ goods gives
\[
 D R_i=\sum_{t=0}^{T-1}\frac{1}{1+x_i(t)}.
\]

We re-write $L_T$ as a product:
\[
 L_T=\sum_{t=0}^{T-1}\frac{1}{n+t}
 =\sum_{t=0}^{T-1}
 \frac{1}{\sqrt{1+x_i(t)}}\cdot\frac{\sqrt{1+x_i(t)}}{n+t}.
\]
By Cauchy--Schwarz, 
\[
 L_T^2\le
 \left(\sum_{t=0}^{T-1}\frac{1}{1+x_i(t)}\right)
 \left(\sum_{t=0}^{T-1}\frac{1+x_i(t)}{(n+t)^2}\right).
\]
The first sum is $D R_i$. Substituting and dividing both sides by
$R_i L_T^2$ gives
\[
 \frac{1}{R_i}\le \frac{D}{L_T^2}
 \sum_{t=0}^{T-1}\frac{1+x_i(t)}{(n+t)^2}.
\]
Now sum over the agents. After $t$ goods have been allocated,
$\sum_{i\in N}x_i(t)=t$, so $\sum_{i\in N}(1+x_i(t))=n+t$. Hence
\begin{eqnarray}
 \sum_{i\in N}\frac{1}{R_i}
 \le \frac{D}{L_T^2}\sum_{t=0}^{T-1}
 \frac{\overbrace{\sum_{i\in N}(1+x_i(t))}^{n+t}}{(n+t)^2}
 =\frac{D}{L_T^2}\underbrace{\sum_{t=0}^{T-1}\frac{1}{n+t}}_{L_T}
 =\frac{D}{L_T}. \label{eq:ub_inverse_Ris}
\end{eqnarray}

Our choice $D=H_nL_T-1$ gives $D/L_T=H_n/(1+1/D)$.
Order the agents so that $R_1\le\cdots\le R_n$.
We claim that $R_h\ge h(1+1/D)$ for some $h$. Suppose instead that
$R_j<j(1+1/D)$ for every $j=1,\ldots,n$.  Summing the $1/R_j$'s 
\begin{eqnarray}\label{eq:rank-threshold}
 \sum_{j=1}^n \frac{1}{R_j}
 > \sum_{j=1}^n \frac{1}{j(1+1/D)}
 = \frac{1}{1+1/D}\sum_{j=1}^n \frac{1}{j}
 = \frac{H_n}{1+1/D},
\end{eqnarray}
contradicting Eq.~\eqref{eq:ub_inverse_Ris}. Thus there is an $h$ for which $R_h\ge h(1+1/D)$. Since the agents
are ordered by increasing $R_i$, every $i\ge h$ also satisfies
$R_i\ge R_h\ge h(1+1/D)$.
Each agent $i\ge h$ must have filled at least $h$ bundles.
Indeed, while any bundle is unfilled, each full bundle accounts for
at most $1+1/D$ in tentative value, since only its last good could
have been reduced. The unfilled bundles together account for less
than $1$, since only the first can contain goods and none of its goods
was reduced. With fewer than $h$ full bundles this gives
$R_i<(h-1)(1+1/D)+1\le h(1+1/D)$, contradicting $R_i\ge R_h$. Thus agents $h,\ldots,n$ have the required number of full bundles.
\end{proof}

We now conclude by giving a tight approximation ratio for the item-by-item case.

\begin{theorem}\label{thm:item-by-item-bound}
Under item-by-item arrivals, no online algorithm can guarantee an
ex post MMS approximation strictly greater than $1/H_n$, even when
it knows in advance that $\mu_i=1$ and $v_i(M)=n$ for every agent $i$.
\end{theorem}

\begin{proof}
For each agent $i$, adding the completion good $g_j$ to their
$j$th auxiliary bundle gives a partition of all goods into $n$
bundles of value exactly $1$. Thus $v_i(M)=n$ and $\mu_i\ge1$.
Since the least valuable bundle in any $n$-partition is worth
at most $v_i(M)/n=1$, we have $\mu_i=1$.

By \Cref{lem:switch}, the adversary starts the second phase
after at most $T$ goods. Fix $n-h+1$ agents who each have at least $h$ full
bundles at that point. Since their bundles were filled in order, all
these agents value $g_1,\ldots,g_h$ at zero, which is the reeamining capacity of these bundles. Only $n-h$ goods remain,
so some agent $i$ among them gets zero value in the second phase.

Consider the $r$th good $g$ allocated to this agent in the first phase.
Just before $g$ arrives, $i$ has been allocated $r-1$ goods, so
$v_i(g)\le q_i(g)=1/(D\cdot r)$. Summing over the goods allocated to
$i$ gives
\[
 v_i(A_i)\le \sum_{r=1}^T \frac{1}{D\cdot r}
 =\frac{H_T}{H_nL_T-1}.
\]
We have that
\[
 H_T-L_T=H_{n-1}-\sum_{j=T+1}^{T+n-1}\frac1j,
 \qquad 0\le H_T-L_T\le H_{n-1},
\]
where $H_0=0$. For fixed $n$, this difference is bounded while
$H_T\to\infty$, since we can choose $T$ arbitrarily large. Dividing by $H_T$ gives $1-L_T/H_T\to0$, so
$L_T/H_T\to1$. Dividing the numerator and denominator of our bound
by $H_T$ now gives
\[
v_i(A_i)\le \frac{H_T}{H_nL_T-1}
 =\frac{1}{H_n(L_T/H_T)-1/H_T}\longrightarrow\frac1{H_n}.
\]

\end{proof}

\subsubsection{Batched arrivals}\label{sec:harmonic-batch-impossibility}

\begin{theorem}\label{thm:batched-bound}
For every $\varepsilon>0$ and all sufficiently large
$\tilde k=\min\{k,n\}$, no online algorithm can guarantee an ex post
MMS approximation of $(1+\varepsilon)/\ln\tilde k$ under $k$-batch
arrivals against an adaptive adversary, even when it knows in advance
that $\mu_i=1$ and $v_i(M)=n$ for every agent $i$.
\end{theorem}

With $k$ batches known in advance, we have only $T=k-2$ batches
between the initial batch $B_1$ and the final batch $B_k$. We use
$B_1$ to reduce the number of agents for whom we need to fill bundles.
Fix $2\le n'<n$, to be chosen below. In $B_1$, introduce $n-n'$ goods worth
$1$ each to every agent. At most $n-n'$ agents receive a good, so
we can choose a set $S$ of $n'$ agents who receive nothing. For each
of them, the initial goods already provide $n-n'$ singleton bundles
of value $1$. It remains to fill $n'$ $1$-valued bundles for each agent in $S$.

As before, we introduce one small good at a time. To reach the second phase
within $T$ such goods, we change how the adversary sets their tentative
values. Set $D=\sqrt{T H_{n'}/n'}-1$, which will be greater than $1$
for our choice of $n'$.
Each agent $i\in S$ starts with weight $w_i=1$. Before introducing the
next first-phase good $g$, let
\[
 \delta=\frac{\sum_{j\in S} w_j}{n'D},
 \qquad q_i(g)=\frac{\delta}{D(w_i+\delta)}.
\]
If $i\in S$ receives $g$, increase $w_i$ by $\delta$. A larger
weight gives an agent a smaller tentative value for a given $\delta$.
As in the item-by-item construction, we cap $v_i(g)$ at the amount
missing from their first unfilled bundle, and set it to zero once all
their bundles are full. Agents outside $S$ value every small good at zero.

Once $n'-h+1$ agents in $S$ each have at least $h$ full bundles for
some $1\le h\le n'$, we reserve the completion goods for $B_k$.
Any remaining intermediate batch contains one good worth zero to everyone.
In $B_k$, introduce $n'$ goods together. To each $i\in S$, the $j$th good
is worth the amount missing from their $j$th bundle. Each agent outside
$S$ values every good in $B_k$ at $1$.

\paragraph{Proof of the batched bound}

If the completion condition is met before the $T$th intermediate batch, there is nothing to prove for this step. Otherwise, all $T$ intermediate batches contain goods generated by the tentative value rule. For each such good $g$, the definitions of $q_i(g)$ and $\delta$ give
\[
    \sum_{i\in S}\frac{1}{q_i(g)}
    =D\left(\frac{\sum_{i\in S}w_i}{\delta}+n'\right)
    =n'D(D+1).
\]
Let $g_1,\ldots,g_T$ be the $T=k-2$ intermediate goods,
one from each intermediate batch. For each $i\in S$, let
$R_i=\sum_{t=1}^T q_i(g_t)$ be their total tentative value for
these goods. Since $q_i(g_t)>0$ for every $t$, Cauchy--Schwarz gives
\[
    T^2
    =\left(\sum_{t=1}^T
        \sqrt{q_i(g_t)}\,\frac{1}{\sqrt{q_i(g_t)}}\right)^2
    \le R_i\sum_{t=1}^T\frac{1}{q_i(g_t)}.
\]
Dividing by $T^2R_i$, summing over $i\in S$, and interchanging
the two sums, we obtain
\[
    \sum_{i\in S}\frac{1}{R_i}
    \le \frac{1}{T^2}\sum_{t=1}^T\sum_{i\in S}
        \frac{1}{q_i(g_t)}
    = \frac{n'D(D+1)}{T}
    = \frac{H_{n'}}{1+1/D},
\]
where the last equality follows from $n'(D+1)^2=T H_{n'}$.
Order the agents in $S$ so that $R_1\le\cdots\le R_{n'}$.
Applying the argument of \eqref{eq:rank-threshold} with $n'$
in place of $n$, the preceding bound implies that
$R_h\ge h(1+1/D)$ for some rank $h$.
By the bundle-filling argument in the proof of
Lemma~\ref{lem:switch}, each of the last $n'-h+1$ agents
has at least $h$ full bundles. Thus the completion condition
holds by the end of $B_{k-1}$.
Note that as in the item-by-item proof, the $n'-h+1$ agents with at least $h$ full bundles all
value the first $h$ goods in $B_k$ at zero. There are only $n'-h$ other goods
in $B_k$, so at least one of these agents receives zero value
from the final batch. This agent also received nothing in $B_1$.
It therefore suffices to bound the value that each selected
agent receives from the intermediate goods.

Fix an agent $i\in S$. If they receive a small good $g$, their weight increases from $w_i$ to $w_i+\delta$. Since
$1/u\ge 1/(w_i+\delta)$ for $w_i\le u\le w_i+\delta$, we have
\[
    v_i(g)
    \le \frac{1}{D}\frac{\delta}{w_i+\delta}
    \le \frac{1}{D}\int_{w_i}^{w_i+\delta}\frac{du}{u}
    = \frac{1}{D}\bigl(\ln(w_i+\delta)-\ln w_i\bigr).
\]

Suppose $i$ receives $s$ small goods, and let $w_i^{(r)}$
be their weight just after receiving the $r$th such good,
with $w_i^{(0)}=1$. Their weight does not change between
these allocations, so just before receiving their $r$th
small good, their weight is $w_i^{(r-1)}$. The bound above
therefore gives a total value of at most
\[
    \frac{1}{D}\sum_{r=1}^{s}
        \bigl(\ln w_i^{(r)}-\ln w_i^{(r-1)}\bigr)
    =\frac{\ln w_i^{(s)}-\ln w_i^{(0)}}{D}
    =\frac{\ln w_i^{(s)}}{D}.
\]
Any padding goods have value zero and contribute nothing
to this sum.

To bound the final weight $w_i^{(s)}$, consider the total
weight $W=\sum_{j\in S}w_j$. Its initial value is $n'$. When a small
good goes to an agent in $S$, $W$ increases by
$\delta=W/(n'D)$, so it is multiplied by $1+1/(n'D)$.
If the recipient is outside $S$, $W$ stays the same.
Since there are at most $T$ small goods, we obtain
\[
    w_i^{(s)}\le n'\left(1+\frac{1}{n'D}\right)^T.
\]
Using $\ln(1+x)\le x$ for $x\ge0$ gives
\[
    \ln w_i^{(s)}\le \ln n'+\frac{T}{n'D}.
\]
Thus, for the agent $i \in S$ who received zero value in
$B_1$ and $B_k$, we have
\[
\begin{aligned}
    v_i(A_i)
    &\le \frac{\ln n'}{D}+\frac{T}{n'D^2}\\
    &=\frac{\ln n'}{D}+\frac{(1+1/D)^2}{H_{n'}},
\end{aligned}
\]
where the equality follows from $T H_{n'}=n'(D+1)^2$.

We now choose
\[
    n'=\left\lfloor\frac{\tilde k}{(\ln\tilde k)^6}\right\rfloor.
\]
For sufficiently large $\tilde k$, this gives $2\le n'<n$.
Also, since $\tilde k/(\ln\tilde k)^6\to\infty$,
\[
    \ln n'=\ln\tilde k-6\ln\ln\tilde k+o(1).
\]
Since $H_t=\ln t+O(1)$, it follows that
$H_{n'}\sim H_{\tilde k}$ as $\tilde k\to\infty$.

To bound $D$, use $T=k-2\ge\tilde k/2$ for sufficiently
large $\tilde k$, along with $H_{n'}\ge1$ and
$n'\le\tilde k/(\ln\tilde k)^6$. These give
\[
    D+1=\sqrt{\frac{T H_{n'}}{n'}}
    \ge\frac{(\ln\tilde k)^3}{\sqrt{2}}.
\]
In particular, $D>1$ for sufficiently large $\tilde k$.
Since $\ln n'\le\ln\tilde k$, we have
$\ln n'/D=O(1/(\ln\tilde k)^2)=o(1/\ln\tilde k)$.
The second term, $(1+1/D)^2/H_{n'}$, is
$(1+o(1))/H_{\tilde k}$. Hence
\[
    v_i(A_i)\le\frac{1+o(1)}{H_{\tilde k}}.
\]

Finally, for each agent in $S$, the $n-n'$ initial goods
form singleton bundles of value $1$, and the final goods
bring each of their $n'$ auxiliary bundles to value $1$.
For each agent outside $S$, the $n-n'$ initial goods and
the $n'$ final goods all have value $1$, and every
intermediate good has value zero. Thus every agent has
a partition of all goods into $n$ bundles worth exactly
$1$ each. It follows that $v_i(M)=n$ and $\mu_i\ge1$,
while $\mu_i\le v_i(M)/n=1$ by averaging. Hence $\mu_i=1$
for every agent, which proves the claimed approximation bound.

Both constructions use only allocations already made, so the
bounds also apply to ex post guarantees for randomized algorithms
against an adaptive adversary.

\subsubsection{A Sharper Bound for Two Agents}
\label{sec:impossibility-2}

\begin{theorem}\label{thm:supp-two}
For two agents, no deterministic online algorithm can guarantee an MMS
approximation strictly greater than $12/19$, even when both MMS values
are known in advance and equal to $1$.
\end{theorem}

\begin{proof}
We describe the construction using integer values and show that both
MMS values are exactly $19$, whereas some agent receives value at most
$12$.  Dividing all values by $19$ gives the stated result.

A \emph{private padding item} for agent $i$ has value $1$ to agent $i$
and value $0$ to the other agent.  When we specify padding $(p_1,p_2)$,
we append $p_i$ such items for agent $i$.  Thus padding can increase
agent $i$'s allocated value by at most $p_i$.

The first item has values $(6,6)$.  Assume without loss of generality
that it is allocated to agent~$1$.  The second item has values $(6,8)$.

First suppose that the second item is also allocated to agent~$1$.
Reveal an item with values $(19,12)$.  If it is allocated to agent~$1$,
append padding $(7,12)$ and stop.  Each agent has total value $38$ and
can partition the items into two bundles of value $19$: agent~$1$ uses
the singleton of value $19$, and agent~$2$ uses the values $12,6,1$.
Agent~$2$ receives value at most $12$.

Otherwise, both agents now have allocated value $12$.  Reveal one final
item with values $(13,13)$.  The resulting valuation rows are
\[
(6,6,19,13)
\qquad\text{and}\qquad
(6,8,12,13).
\]
The first row has MMS $19$: the subset with values $6,13$ has value
$19$, and there is no subset of value between $20$ and $24$.  The
second row has total value $39$ and a subset of value $19$.  Hence its
MMS is also $19$.  One agent receives none of the final item and
retains value $12$.

We may therefore assume that the second item is allocated to agent~$2$.
Reveal a third item with values $(2,2)$.

\paragraph{Case 1: the third item is allocated to agent~$1$.}
The agents' allocated values are $(8,8)$.  Reveal up to four items,
each with values $(1,4)$, stopping this sequence when the first such
item is allocated to agent~$2$.

Suppose that this happens at its $t$-th item, where $1\le t\le4$.
The allocated values are then $(7+t,12)$.  Reveal one item with values
$(19,23-4t)$, append padding $(5-t,0)$, and stop.  If the last
nonpadding item goes to agent~$1$, agent~$2$ retains $12$.  If it goes
to agent~$2$, agent~$1$ receives value at most
\[
(7+t)+(5-t)=12.
\]
Agent~$1$ has total value $38$ and a singleton of value $19$.
Agent~$2$ has total value $39$, and a subset of value $19$ is given by
\[
\begin{array}{c|cccc}
t&1&2&3&4\\ \hline
\text{subset values}&19&15+4&11+8&7+8+4.
\end{array}
\]
Thus both MMS values are exactly $19$.

It remains to consider the case where all four $(1,4)$-items go to
agent~$1$.  The allocated values are $(12,8)$.  Reveal an item with
values $(19,4)$.  If it goes to agent~$1$, append padding $(1,2)$ and
stop.  Both total values are $38$.  Agent~$1$ has a singleton of value
$19$, and agent~$2$ has a subset with values $8,6,4,1$, also worth
$19$.  Agent~$2$ receives value at most $8+2=10$.

If the $(19,4)$-item goes to agent~$2$, both allocated values become
$12$.  Reveal one last item with values $(2,3)$ and stop.  Both total
values are now $39$.  Agent~$1$ has a singleton of value $19$, and
agent~$2$ has a subset with values $8,4,4,3$, worth $19$.  Both MMS
values are exactly $19$, and one agent retains value $12$.

\paragraph{Case 2: the third item is allocated to agent~$2$.}
The allocated values are $(6,10)$.  Reveal up to two items, each with
values $(6,1)$, stopping this sequence when the first such item goes to
agent~$1$.

If the first $(6,1)$-item goes to agent~$1$, the allocated values are
$(12,10)$.  Reveal an item with values $(19,19)$, append padding $(0,2)$,
and stop.  Agent~$1$ has total value $39$ and a singleton of value $19$;
agent~$2$ has total value $38$ and a singleton of value $19$.  Whichever
agent does not receive the $(19,19)$-item finishes with value at most
$12$.

If the first $(6,1)$-item goes to agent~$2$ and the second goes to
agent~$1$, the allocated values are $(12,11)$.  Reveal an item with
values $(13,19)$, append padding $(0,1)$, and stop.  Agent~$1$ has total
value $39$ and a subset with values $13,6$, worth $19$; agent~$2$ has
total value $38$ and a singleton of value $19$.  Again, the agent who
does not receive the last nonpadding item finishes with value at most
$12$.

Finally, suppose that both $(6,1)$-items go to agent~$2$.  The allocated
values are $(6,12)$.  Reveal an item with values $(6,19)$.  If it goes
to agent~$2$, append padding $(6,1)$ and stop.  Agent~$1$ receives value
at most $12$.  Both total values are $38$; agent~$1$ has a subset with
values $6,6,6,1$, worth $19$, and agent~$2$ has a singleton of value
$19$.

If the $(6,19)$-item goes to agent~$1$, both allocated values become
$12$.  Reveal one last item with values $(7,2)$ and stop.  Both total
values are $39$; agent~$1$ has a subset with values $6,6,7$, worth
$19$, and agent~$2$ has a singleton of value $19$.  Both MMS values are
exactly $19$, and one agent retains value $12$.
\end{proof}



\subsection{Two Batches}

We study the two-batch model. We show that no online algorithm can
achieve a $2/3$-approximation to the MMS, in contrast to the setting
of identical valuations. Our impossibility result reduces to a finite
three-agent hard instance. On the positive side, we give a
$4/7$-approximation algorithm, using local exchange properties to show
that no agent remains uncovered.

\subsubsection{$4/7$-approximation Algorithm}

Let $\rho_{+}=4/7$.
An agent is \emph{covered} once their bundle has value at least
$\rho_{+}$. Covered agents receive no further items while
uncovered agents remain. Among the uncovered agents, at most
one has a nonempty bundle. This agent is the \emph{residual agent}.

For a batch $B$, let $U$ be the set of uncovered agents and $P_i$
the current bundle of each $i\in U$. A feasible extension consists
of a set $C\subseteq U$ and pairwise disjoint sets $S_i\subseteq B$
for $i\in C$, satisfying
\[
v_i(P_i\cup S_i)\ge\rho_{+}\qquad(i\in C).
\]
If $U\setminus C\neq\emptyset$, let
$R=B\setminus\bigcup_{i\in C}S_i$ and assign all of $R$ to some
$r\in U\setminus C$ satisfying
$v_r(P_r\cup R)<\rho_{+}$.
If the old residual agent is still uncovered, that agent must
remain the residual agent. If $U\setminus C=\emptyset$, any items
not used in completions may be allocated arbitrarily.

Among all feasible extensions, choose one lexicographically by:
\begin{enumerate}
    \item maximizing $|C|$;
    \item maximizing the number of singleton completions
    $S_i=\{g\}$ with $v_i(g)\ge\rho_{+}$;
    \item minimizing such singleton completions for agents with
    $P_i\neq\emptyset$;
    \item minimizing $\sum_{i\in C}|S_i|$;
    \item lexicographically maximizing the nondecreasing vector
    $(v_i(S_i):i\in C)$;
    \item if an uncovered agent remains, maximizing its resulting
    residual value.
\end{enumerate}
A feasible extension always exists: assign the whole batch to the
old residual agent, or to any uncovered agent if none exists.
If that agent reaches $\rho_{+}$, cover it; otherwise it remains
residual. The invariant is thus maintained after each batch.

The lexicographic choice yields the following local exchange bounds.

\begin{lemma}\label{lem:two-batch-exchange}
Fix a batch, an agent $u$ who remains uncovered after it, and an
agent $i$ covered during it. If $|S_i|\ge2$, then
\[
v_i(g)<\rho_{+},\qquad v_u(g)<\rho_{+}
\qquad(g\in S_i),
\]
and $v_i(P_i\cup S_i)<2\rho_{+}$.
Moreover, if $|S_i|\ge2$ and $v_u(S_i)\ge\rho_{+}$, then
$v_u(S_i)\le v_i(S_i)$.
If $S_i=\{g\}$ and $v_u(g)\ge\rho_{+}$, then
$P_i=\emptyset$.
Finally, if $|S_i|=t\ge3$, then
\[
v_u(S_i)<\frac{t}{t-1}\rho_{+}\le\frac32\rho_{+}.
\]
\end{lemma}

\begin{proof}
In each exchange below, released current-batch items are assigned
to the appropriate residual agent. If this creates an additional
covered agent, criterion~1 is contradicted; otherwise the modified
allocation remains feasible.

Suppose $|S_i|\ge2$. If $v_i(g)\ge\rho_{+}$ for some $g\in S_i$,
replacing $S_i$ by $\{g\}$ improves criterion~2. If
$v_u(g)\ge\rho_{+}$, use $\{g\}$ to cover $u$ instead of $i$;
this either creates an additional completion or improves
criterion~2. Thus $v_i(g),v_u(g)<\rho_{+}$ for every $g\in S_i$.

Every item of $S_i$ is necessary for covering $i$, since otherwise
removing it improves criterion~4. Hence
\[
v_i(P_i\cup(S_i\setminus\{g\}))<\rho_{+}
\qquad(g\in S_i),
\]
which, together with $v_i(g)<\rho_{+}$, gives
$v_i(P_i\cup S_i)<2\rho_{+}$.

If $v_u(S_i)\ge\rho_{+}$, replace the completion of $i$ by that of
$u$ using the same set $S_i$. Unless an additional completion
becomes possible, the first four criteria are unchanged; criterion~5
therefore implies $v_u(S_i)\le v_i(S_i)$.

Suppose $S_i=\{g\}$, $v_u(g)\ge\rho_{+}$, and
$P_i\neq\emptyset$. The residual invariant implies
$P_u=\emptyset$. Assigning $g$ to $u$ instead either increases
the number of direct singleton completions or leaves that number
unchanged and improves criterion~3. Thus $P_i=\emptyset$.

Finally, let $|S_i|=t\ge3$. For every $g\in S_i$,
\[
v_u(S_i\setminus\{g\})<\rho_{+},
\]
since otherwise $u$ could be covered with fewer current-batch
items, contradicting criterion~4 (or criterion~1 if an additional
completion becomes possible). Summing these $t$ inequalities yields
$(t-1)v_u(S_i)<t\rho_{+}$, proving the final claim.
\end{proof}

The next lemma controls the agent carrying a nonempty first-batch
residual into the second batch.

\begin{lemma}\label{lem:two-batch-residual}
Let $P$ be a nonempty residual set after the first batch and let
$r$ be its recipient. For every agent $j$ who remains uncovered
after that batch,
\[
v_r(P)\ge v_j(P).
\]
Moreover, if $r$ is covered in the second batch, then for every
agent $u$ who remains uncovered at the end,
$v_u(A_r)<2\rho_{+}$, where $A_r$ is $r$'s final bundle.
\end{lemma}

\begin{proof}
Every agent $j$ remaining uncovered after the first batch satisfies
$v_j(P)<\rho_{+}$; otherwise the same first-batch completions could
be kept and $P$ assigned to $j$, creating an additional covered
agent. Criterion~6 therefore chooses $r$ so that
$v_r(P)\ge v_j(P)$.

Suppose $r$ is later covered using $S_r$.
If $S_r$ is a singleton, \Cref{lem:two-batch-exchange} and
$P_r=P\neq\emptyset$ imply $v_u(S_r)<\rho_{+}$, while
$v_u(P)<\rho_{+}$. The same conclusion holds when
$|S_r|\ge2$ and $v_u(S_r)<\rho_{+}$.

Otherwise $v_u(S_r)\ge\rho_{+}$. Then
\Cref{lem:two-batch-exchange} and the first part give
\[
v_u(P)+v_u(S_r)
\le v_r(P)+v_r(S_r)<2\rho_{+},
\]
where the final inequality follows from
\Cref{lem:two-batch-exchange} applied to $r$.
\end{proof}

\begin{theorem}\label{thm:two-batch-possibility}
The algorithm guarantees a $4/7$-approximation to the MMS.
\end{theorem}

\begin{proof}
Suppose some agent remains uncovered after the second batch, and
let $u$ be the final residual agent.

Partition the covered agents into four groups:
\begin{itemize}
    \item $h$ agents whose final bundles are singletons;
    \item $q$ agents whose final bundles consist of two items from
    one batch;
    \item $\ell$ agents whose final bundles consist of at least
    three items from one batch;
    \item $e$ agents who held a nonempty first-batch residual
    bundle and were covered in the second batch.
\end{itemize}
Since there is only one first-batch residual agent, $e\le1$.
Let $a\ge1$ be the number of agents still uncovered at the end.
Thus $n=h+q+\ell+e+a$.

Take an MMS partition of $u$ into $n$ bundles of value at least $1$.
Remove every part containing one of the $h$ items allocated as a
singleton bundle. This removes at most $h$ parts; if fewer are
removed, discard additional parts so that exactly
$N=n-h=q+\ell+e+a$ parts remain.

Let $T$ be the set of the $2q$ items allocated to the two-item
recipients. By \Cref{lem:two-batch-exchange}, every item of $T$ is
worth less than $\rho_{+}$ to $u$. If a retained MMS part $Q$
contains $t$ items of $T$, then
\[
v_u(Q\setminus T)\ge(2-t)(1-\rho_{+}).
\]
For $t=0$, this follows from
$v_u(Q)\ge1\ge2(1-\rho_{+})$;
for $t=1$, deleting an item worth less than $\rho_{+}$ leaves
value greater than $1-\rho_{+}$; for $t\ge2$ the right-hand
side is nonpositive.

Let $W$ be the union of the retained parts after removing $T$.
These parts contain at most $2q$ items of $T$. Summing the preceding
inequality therefore yields
\[
v_u(W)\ge2(N-q)(1-\rho_{+})
=2(\ell+e+a)(1-\rho_{+}).
\]
Every item of $W$ is allocated either to an agent in the
$\ell$-group, to the possible cross-batch residual agent, or to
a final uncovered agent. By \Cref{lem:two-batch-exchange}, every
$\ell$-agent contributes less than $3\rho_{+}/2$ to $u$.
By \Cref{lem:two-batch-residual}, the possible cross-batch
residual agent contributes less than $2\rho_{+}$.
Finally, all final uncovered agents together contribute less than
$\rho_{+}$, since only $u$ has a nonempty bundle. Hence
\[
2(\ell+e+a)(1-\rho_{+})
<\frac32\rho_{+}\ell+2\rho_{+}e+\rho_{+}.
\]
Substituting $\rho_{+}=4/7$ gives
\[
\frac67(\ell+e+a)<\frac67\ell+\frac87e+\frac47,
\]
and therefore $6a-2e<4$, a contradiction to $a\ge1$ and
$e\le1$. Thus every agent is covered and receives at least
$4/7$ of their MMS.
\end{proof}

\subsubsection{$<2/3$-Inapproximability}

\begin{theorem}\label{thm:two-batch-impossibility}
For every $n\ge6$, no online algorithm for two-batch arrivals with exact
MMS advice can guarantee ex post an approximation strictly greater than
$\frac23-\frac1{2322429}$.
\end{theorem}

We first establish a reduction from a finite hard instance to the
two-batch setting.

\begin{lemma}\label{lem:two-batch-reduction}
Suppose that a finite set $G$ admits $m$ nonnegative integer-valued
additive valuations $w_1,\ldots,w_m$ and integers $H\le L<2H$
such that:
\begin{enumerate}
    \item for every $i$, $w_i(G)=2mH$, and $G$ has a partition into
    $2m$ bundles of value exactly $H$ to agent $i$;
    \item every allocation of $G$ to the $m$ agents gives some agent
    value at most $L$.
\end{enumerate}
Then, for every $n\ge2m$, no deterministic two-batch algorithm with
exact MMS advice can guarantee more than $L/(H+L)$.
\end{lemma}

\begin{proof}
Let $\rho_{-}=L/(H+L)\in[1/2,2/3)$.
The first batch contains $n-2m$ unit-valued items and $2m$ items of
value $\rho_{-}$, identically valued by all agents.

Let $x$ be the number of agents whose first-batch value is strictly
larger than $\rho_{-}$. At most $n-2m$ of them receive a unit item,
while every other such agent needs at least two $\rho_{-}$-items.
Hence $x\le n-m$.

If $x<n-m$, reveal $m$ unit items. More than $m$ agents had value
at most $\rho_{-}$, so one of them receives no new item and finishes
with value at most $\rho_{-}$. The MMS is exactly $1$: the $n-m$
unit items and $m$ pairs of $\rho_{-}$-items give $n$ bundles of
value at least $1$. Conversely, there are fewer than $2n$ positive
items, all worth at most $1$, so every $n$-partition has a bundle
containing at most one positive item, of value at most $1$.

Suppose instead that $x=n-m$. Equality forces all $n-2m$ unit items
to be assigned to distinct agents and all $2m$ $\rho_{-}$-items in
pairs to $m$ further agents. Thus exactly $m$ agents receive nothing.

For these $m$ agents, reveal $G$ with valuations $w_i/(H+L)$; give
every other agent any one of these valuations on $G$. Every agent
has total value
\[
(n-2m)+2m\rho_{-}+\frac{2mH}{H+L}=n.
\]
Furthermore, the $n-2m$ unit items and the $2m$ bundles obtained by
pairing each $\rho_{-}$-item with a value-$H$ part of $G$ form an
$n$-partition of value exactly $1$. Hence every MMS is exactly $1$.

If all $m$ initially empty agents received more than $\rho_{-}$
from $G$, assigning to them also any items of $G$ allocated to other
agents would yield an allocation of $G$ giving every one of them
more than $L$, contrary to assumption~2. Thus some agent receives
at most $\rho_{-}$.
\end{proof}

We use the following explicit three-agent instance. Its item data
are listed in \Cref{app:two-batch-impossibility}; the argument below gives
the complete verification of the properties needed by the reduction.

\begin{lemma}\label{lem:two-batch-certificate}
There exists a $24$-item instance with three additive integer-valued
valuations satisfying \Cref{lem:two-batch-reduction} with
$m=3$, $H=258048$, and $L=516095$.
\end{lemma}

\begin{proof}
Let $G=\{0,\ldots,23\}$. For each item $e$, the supplementary table
specifies a positive integer base value $b_e$, perturbations
$r_e,c_e,z_e$, and labels $i_e,j_e,h_e\in\{0,\ldots,5\}$.
Define
\[
    w_R(e)=7b_e+r_e,\qquad
    w_C(e)=7b_e+c_e,\qquad
    w_Z(e)=7b_e+z_e.
\]
For every $t\in\{0,\ldots,5\}$, the items with $i_e=t$ have total
base value $36864$ and total $r$-perturbation zero. Similarly,
the $j$-classes have base value $36864$ and zero
$c$-perturbation, and the $h$-classes have base value $36864$
and zero $z$-perturbation. Thus each valuation has a partition into
six bundles worth $H=7\cdot36864=258048$. Consequently every agent
has total value $6H$. All item values are positive.

It remains to prove that no allocation gives all three agents
value at least $2H=516096$. The sums of the positive perturbations
of $r,c,z$ are $7,6,8$, respectively. Hence any bundle reaching
$516096$ has base value at least
\[
    \left\lceil\frac{516096-8}{7}\right\rceil=73727.
\]
The total base value is $6\cdot36864=221184$. Therefore, if all
three bundles reach $516096$, each must also have base value at most
$221184-2\cdot73727=73730$.

For $X\in\{R,C,Z\}$, define
\[
\mathcal F_X=\left\{S\subseteq G:
73727\le b(S)\le73730,\;w_X(S)\ge516096\right\}.
\]
An exact exhaustive enumeration of the $2^{24}$ subsets of $G$
gives
\[
|\mathcal F_R|=287,\qquad
|\mathcal F_C|=292,\qquad
|\mathcal F_Z|=296,
\]
and
\[
\left\{(S,T)\in\mathcal F_R\times\mathcal F_C:
S\cap T=\emptyset,\;
G\setminus(S\cup T)\in\mathcal F_Z\right\}=\emptyset.
\]
This verification is exhaustive: every possible bundle is a subset
of $G$, and each family contains exactly those subsets that could
serve as the corresponding agent's bundle in an allocation giving
all three agents at least $516096$. Such an allocation would produce
a pair $(S,T)$ in the displayed empty set. All computations use
integers, with no numerical approximation. Therefore every
allocation gives some agent value at most $L=516095$.
\end{proof}

\begin{proof}[Proof of \Cref{thm:two-batch-impossibility}]
Apply \Cref{lem:two-batch-reduction} to the instance of
\Cref{lem:two-batch-certificate}. Since
\[
\rho_{-}=\frac{L}{H+L}
=\frac{516095}{774143}
=\frac23-\frac1{2322429},
\]
the claimed bound follows. The adversary selects the second batch
after observing the first-batch allocation, so the argument applies
pathwise to any algorithm claiming an ex-post guarantee, including
one that randomizes its allocation.
\end{proof}


\section{Identical  Valuations} \label{sec:identical}


\subsection{A Tight $2/3$-MMS Approximation for Two Batches} \label{sec:results-identical-2batches}

We begin with two batches, where the optimal guarantee can be determined
exactly: a \(2/3\)-approximation is achievable, and no better guarantee is
possible.

\begin{lemma}
   For two batches, no approximation strictly better than $2/3$ is achievable
for any $n\ge2$.
\end{lemma}
\begin{proof}
We adapt the two-agent lower-bound construction of \citet[Theorem~2.2]{online-covering97}.
The first batch consists of $n-2$ items of value $1$ and two items of
value $2/3$.

If the two $2/3$-items are allocated to different agents, then after the
first batch at most $n-2$ agents have value strictly greater than $2/3$,
since every such agent must receive one of the $n-2$ unit-valued items.
The second batch then consists of one item of value $1$, so some agent
still receives value at most $2/3$.

If the two $2/3$-items are allocated to the same agent, then after the
first batch some agent has value $0$.  The second batch then consists of
two items of value $1/3$, whose total value is $2/3$, so this agent can
receive value at most $2/3$.

In both cases the final instance has MMS exactly $1$.  In the first case,
the $n-1$ unit-valued items together with the two $2/3$-items admit a
partition into $n$ bundles of value at least $1$, while a partition into
$n$ bundles of value strictly greater than $1$ is impossible.  In the
second case, the $n-2$ unit-valued items and the two pairs consisting of
a $2/3$-item and a $1/3$-item form an MMS partition of value $1$, and the
total value is exactly $n$.
\end{proof}
\paragraph{A $2/3$ Approximation Algorithm.}
An agent is \emph{active} while the value of their current allocation is
less than $2/3$, and \emph{inactive} otherwise.
For a batch $B$, let $\kappa(B)$ be the maximum number, capped at $n$, of pairwise
disjoint bundles of value at least $2/3$ that can be formed from $B$.
Among all such maximum collections, let $\rho(B)$ be the maximum value
of the remaining items.

In $B_1$, choose a collection attaining both $\kappa(B_1)$ and $\rho(B_1)$, allocate its bundles to distinct agents, and, if active agents remain, give all residual items to one designated residual agent (if $\kappa(B_1)=n$, all players are inactive, and the remaining items are assigned arbitrarily).
In $B_2$, compute an analogous collection.  If $\kappa(B_1)+\kappa(B_2)\ge n$, allocate enough of its bundles to the
remaining agents.  Otherwise, allocate them to active agents other
than the residual agent and give the residual items of $B_2$ to that
agent.

Since valuations are identical, only the number of $2/3$-bundles matters,
and the agents receiving them can always be chosen as above.  Moreover,
whenever $\kappa(B)<n$, we have $\rho(B)<2/3$, since otherwise the
residual itself would form one additional $2/3$-bundle.

\begin{theorem}
    Our algorithm guarantees a $2/3$-approximation to the MMS.
\end{theorem}

\begin{proof}
Let $\{Q_1,\ldots,Q_n\}$ be an MMS partition, so $v(Q_i)\ge1$ for every
$i$.  Write
$b_{j,i}=v(Q_i\cap B_j)$ for $j\in\{1,2\}$, and classify each $Q_i$ as
$B_1$-heavy if $b_{1,i}\ge2/3$, as $B_2$-heavy if
$b_{1,i}<2/3$ and $b_{2,i}\ge2/3$, and as mixed otherwise.
Let $a,b,c$ denote the numbers of bundles of these three types, so
$a+b+c=n$.

Every $B_1$-heavy bundle certifies a $2/3$-bundle in $B_1$, and every
$B_2$-heavy bundle certifies one in $B_2$.  Moreover, for every mixed
bundle,
$b_{1,i},b_{2,i}>1/3$.
Hence each pair of mixed bundles certifies one additional $2/3$-bundle
in each batch.  Therefore
\[
\kappa(B_1)\ge a+\left\lfloor\frac c2\right\rfloor,
\qquad
\kappa(B_2)\ge b+\left\lfloor\frac c2\right\rfloor.
\]

If $c$ is even, their sum is at least $n$, and all agents can be made inactive.
Suppose that $c=2d+1$.  Then
$\kappa(B_1)+\kappa(B_2)\ge n-1$.
If the inequality is strict, we are again done.

Otherwise,
$\kappa(B_1)=a+d$ and $\kappa(B_2)=b+d$.
Pair $2d$ of the mixed bundles and let $Q_\ell$ be the remaining one.
The corresponding heavy bundles and paired mixed bundles form maximum
$2/3$-packings in the two batches, leaving
$Q_\ell\cap B_1$ and $Q_\ell\cap B_2$, respectively, in their residuals.
Thus, by maximality of the residual values,
\[
\rho(B_1)+\rho(B_2)
\ge
v(Q_\ell\cap B_1)+v(Q_\ell\cap B_2)
=
v(Q_\ell)
\ge1.
\]
Hence the residual agent also receives value at least $2/3$, completing
the allocation.
\end{proof}

\subsection{A \texorpdfstring{$(k+1)/(2k+1)$}{(k+1)/(2k+1)}
Approximation for $k$ Batches}
\label{sec:identical-k-batches}

Assume that all agents have the same nonnegative additive
valuation $v$ and that the common MMS is normalized to $1$.
Fix $k\geq3$ and let $\tau=(k+1)/(2k+1)$.
We call an agent \emph{covered} once the value of their
bundle reaches $\tau$.
At any time, the algorithm keeps at most one uncovered
agent with a nonempty bundle; we call this agent the
\emph{residual agent}.

\paragraph{Algorithm.}
When a batch arrives, the algorithm first processes its
items of value at least $\tau$.
Each such item is assigned alone to an empty uncovered
agent.
If no empty uncovered agent remains, then either all
agents are already covered or there is one residual
agent; in the latter case, the next such item is
assigned to the residual agent and covers the last
uncovered agent.
Once all agents are covered, the remaining items may
be allocated arbitrarily.

The algorithm then sorts all remaining items of the
batch in nonincreasing order of value.
It assigns them one by one to the residual agent,
if one exists, and otherwise to an arbitrary empty
uncovered agent.
Whenever the current agent reaches value $\tau$,
they are declared covered and the algorithm continues
with an empty uncovered agent.
If the batch ends before the current agent is covered,
that agent becomes the residual agent for the next batch.
Thus, at the end of every batch, every uncovered agent
is empty except possibly the residual agent.

\begin{theorem}
\label{thm:identical-k-batches}
For identical additive valuations and $k\geq3$ batches,
the algorithm guarantees every agent a value of at least
$(k+1)/(2k+1)$ times their MMS.
The guarantee continues to hold when each batch is
selected after the adversary observes all earlier
allocations.
\end{theorem}

We prove the theorem using the weight-function paradigm
of semi-online machine covering
\citep{semi-online-covering}.
An MMS partition gives a lower bound on the total weight,
whereas an unsuccessful execution gives a conflicting
upper bound.
The batched structure allows the excess weight to be
charged to at most one mixed bundle per batch.

\paragraph{Weight function.}
Let
$\sigma=k/(2k+1)$ and
$\lambda=(2k+1)/(2k)$.
We will use the identities
\[
\tau+\sigma=1,\qquad
\tau-\sigma=\frac{1}{2k+1},\qquad
\lambda\sigma=\frac12,
\]
and
\[
\lambda\tau=\frac{k+1}{2k},
\qquad
\lambda=1+\frac{1}{2k}.
\]

Call an item \emph{large} if its value is at least
$\tau$, \emph{medium} if its value is strictly between
$\sigma$ and $\tau$, and \emph{small} if its value is
at most $\sigma$.
Assign weight $1$ to each large item, $1/2$ to each
medium item, and $\lambda$ times its value to each
small item.
Let $w(S)$ be the sum of the weights of the items
in a set $S$.

We first relate the weight of a bundle to its value.

\begin{lemma}
\label{lem:identical-k-benchmark}
If $v(Q)\geq1$, then $w(Q)\geq1$.
If $v(R)<\tau$, then $w(R)<\lambda\tau$.
\end{lemma}

\begin{proof}
For the first statement, a bundle containing a large
item has weight at least $1$, as does a bundle
containing at least two medium items.

If a bundle contains exactly one medium item of
value $p$, then its small items have total value
at least $1-p>\sigma$, so its weight is greater
than $1/2+\lambda\sigma=1$.
If it contains only small items, its weight is
$\lambda v(Q)\geq\lambda>1$.

For the second statement, $R$ contains no large item
and at most one medium item.
If it contains no medium item, then
$w(R)=\lambda v(R)<\lambda\tau$.
If it contains one medium item of value $p$,
then its small items have total value less than
$\tau-p<\tau-\sigma$, and therefore
\[
w(R)
<\frac12+\lambda(\tau-\sigma)
=\frac12+\frac{1}{2k}
=\lambda\tau.
\]
\end{proof}

For a covered agent, their \emph{closing bundle} is
the set of items they hold when their value first
reaches $\tau$.
A closing bundle is \emph{pure} if its owner was
empty at the beginning of the batch in which it
closed, and \emph{mixed} otherwise.
In an unsuccessful execution, covered agents receive
no later items, so their closing bundles are also
their final bundles.

We next bound the weights of pure and mixed closing
bundles separately.

\begin{lemma}
\label{lem:identical-k-pure}
Every pure closing bundle has weight at most $1$.
\end{lemma}

\begin{proof}
A large item is assigned alone and has weight $1$.
Two medium items already reach the target and have
total weight $1$.

Now suppose that the bundle contains exactly one
medium item of value $p$.
If the first following small item closes the bundle,
the bundle has weight at most
$1/2+\lambda\sigma=1$.

Otherwise, that first small item has value less than
$\tau-p<\tau-\sigma$.
Since the items are processed in nonincreasing order
and the bundle closes as soon as it reaches $\tau$,
all the small items in the closing bundle have total
value less than $2(\tau-p)$.
Its weight is therefore less than
\[
\frac12+2\lambda(\tau-\sigma)
=\frac12+\frac1k
\leq1.
\]

It remains to consider a closing bundle containing
only small items.
If it contains two items, its weight is at most
$2\lambda\sigma=1$.

If it contains at least three items, let $x$ be
its last item.
Immediately before $x$ is added, the bundle has
value less than $\tau$, and at least two earlier
items are no smaller than $x$.
Hence $x<\tau/2$, the final value is less than
$3\tau/2$, and the weight is less than
\[
\frac32\lambda\tau
=\frac{3(k+1)}{4k}
\leq1,
\]
where the last inequality uses $k\geq3$.
\end{proof}

\begin{lemma}
\label{lem:identical-k-mixed}
In any execution in which the algorithm does not
cover all agents, every mixed closing bundle has
weight at most $\lambda$.
\end{lemma}

\begin{proof}
Let $R$ be the residual bundle at the beginning
of the closing batch, let $r=v(R)<\tau$, and let
$T$ be the sequence of current-batch items added
to $R$ until it closes.
The items in $T$ occur in nonincreasing order.

Moreover, $T$ contains no large item: the algorithm
assigns a large item to the residual agent only
when no empty uncovered agent remains, and that
assignment would cover the last uncovered agent.
By \Cref{lem:identical-k-benchmark},
$w(R)<\lambda\tau$.

Suppose first that $T$ contains a medium item $p$.
This item is necessarily the first item of $T$.
If it closes $R$, then the closing bundle has
weight less than
$\lambda\tau+1/2=\lambda$.

Otherwise, $r<\tau-p<\tau-\sigma$.
In particular, $R$ contains only small items and
has weight less than
$\lambda(\tau-\sigma)=1/(2k)$.

If a second medium item follows, the closing bundle
has weight less than
$1/(2k)+1=\lambda$.
Otherwise, only small items follow $p$.
If one small item is enough, its weight is at most
$\lambda\sigma=1/2$, and the total weight is
less than $\lambda$.

If at least two small items are needed, let
$d=\tau-r-p$.
The first small item does not close the bundle,
so by the ordering and minimality of the closing
prefix, the total value of the small items is
less than $2d<2(\tau-\sigma)$.
Their total weight is therefore less than $1/k$,
and the weight of the closing bundle is less than
\[
\frac12+\frac{3}{2k}\leq\lambda.
\]

Suppose next that $T$ contains only small items
but $R$ contains a medium item $p$.
If the first new item closes $R$, then the total
weight is less than
$\lambda\tau+\lambda\sigma=\lambda$.

Otherwise, the total value of the items in $T$
is less than $2(\tau-r)$.
The small items already in $R$ have total value
$r-p$, and hence the closing bundle has weight
less than
\[
\frac12+\lambda(2\tau-r-p).
\]
Since $r\geq p>\sigma$, we have $r+p>2\sigma$,
and the last expression is less than
\[
\frac12+2\lambda(\tau-\sigma)
=\frac12+\frac1k
\leq\lambda.
\]

Finally, suppose that all items in $R\cup T$
are small.
We show that their total value is at most $1$,
which implies that their weight is at most $\lambda$.

If $T$ contains one item, the total value is
less than $\tau+\sigma=1$.

If $T$ contains two items $x_1\geq x_2$,
then $r+x_1<\tau$.
When $r\leq\tau-\sigma$, their total value is
at most
$r+2\sigma\leq\tau+\sigma=1$.
When $r>\tau-\sigma$, it is less than
$r+2(\tau-r)=2\tau-r<\tau+\sigma=1$.

If $T$ contains at least three items and $x$
is its last item, then the two earlier items
are at least as large as $x$, while the value
added before $x$ is less than $\tau-r$.
Thus $2x<\tau-r$, and the final value is less than
\[
\tau+x<\frac32\tau\leq1.
\]
This completes the proof.
\end{proof}

\begin{proof}[Proof of \Cref{thm:identical-k-batches}]
Consider first the normalized case in which the
common MMS is $1$.
An MMS partition consists of $n$ bundles, each
of value at least $1$.
By \Cref{lem:identical-k-benchmark}, each of these
bundles has weight at least $1$, and therefore
$w(M)\geq n$.

Suppose, for contradiction, that the algorithm
covers only $\ell\leq n-1$ agents.

The residual-agent invariant follows directly
by induction over the batches: the algorithm
works on one uncovered agent at a time and
moves to an empty agent only after covering
the current one.
Therefore, at the beginning of every batch
there is at most one nonempty uncovered bundle.

The first closing bundle in a batch may contain
this old residual, but every later closing bundle
in the same batch starts from an empty agent.
The first batch has no old residual, so the
number $h$ of mixed closing bundles is at most
$k-1$.

By \Cref{lem:identical-k-pure,lem:identical-k-mixed},
the covered agents have total weight at most
\[
(\ell-h)+h\lambda
=\ell+h(\lambda-1)
=\ell+\frac{h}{2k}.
\]

Every other item belongs to the final residual
bundle; all remaining uncovered agents are empty.
The final residual has value less than $\tau$,
so \Cref{lem:identical-k-benchmark} gives it
weight less than
$\lambda\tau=(k+1)/(2k)$.

Consequently,
\[
\begin{aligned}
w(M)
&<\ell+\frac{h}{2k}+\frac{k+1}{2k}\\
&\leq\ell+\frac{k-1}{2k}+\frac{k+1}{2k}\\
&=\ell+1\leq n.
\end{aligned}
\]
This contradicts $w(M)\geq n$ and proves that
every agent receives value at least $\tau$.

For an arbitrary positive common MMS, scale
all values by its reciprocal and apply the
normalized result.
If the common MMS is zero, the claimed
multiplicative guarantee is immediate.

Finally, the proof is pointwise: it applies
to every completed sequence of batches that
is consistent with the announced MMS.
It therefore does not require the adversary
to choose all batches in advance.
\end{proof}

\begin{corollary}
\label{cor:combined-identical-k-batches}
For identical valuations and $k\geq3$, one can
guarantee $\max\{6/11,(k+1)/(2k+1)\}$ of the MMS.
In particular, the guarantee is $4/7$ for three
batches, $5/9$ for four batches, and $6/11$
for every $k\geq5$.
\end{corollary}

\begin{proof}
The semi-online machine-covering algorithm of
\citet{semi-online-covering} guarantees $6/11$
of the known optimum under item-by-item arrivals.
With identical valuations, this optimum is the
common MMS, and the algorithm can be simulated
inside each batch by processing its items in
any fixed order.
Choosing the better of that algorithm and the
algorithm above gives the stated bound.
\end{proof}

\subsection{Impossibility Results for Three, Four and Six Batches}\label{sec:impossible-c-batches}
We complement the preceding guarantees with upper bounds on the
fraction of MMS that an online algorithm can guarantee in each
of these three settings.

\begin{lemma}[Three-batch impossibility]
\label{lem:three-batches-twelve-nineteenths}
For identical nonnegative additive valuations and every $n\geq 38$,
no online algorithm with exact MMS advice can guarantee an
approximation strictly greater than $12/19$ in three batches.
\end{lemma}

\begin{proof}
We construct an adaptive adversary such that every terminal instance
has common MMS exactly $19$, while some agent receives value at most
$12$. Dividing all item values by $19$ gives the claimed normalized
construction with exact MMS advice $1$.

All agents have the same additive valuation throughout the
construction. We first consider $38$ agents and then extend the
construction to every $n>38$.

\paragraph{The construction for $38$ agents.}
The first batch consists of $38$ items of value $12$.
Let $x$ be the number of agents receiving at least two of these
items. Since there are only $38$ items, $x\leq 19$.

\medskip
\noindent
\textbf{Case 1: $x\leq 18$.}
The second batch consists of $19$ items of value $19$.
The third batch consists of a single zero-valued item.

After the first batch, exactly the agents receiving at least two
items have value greater than $12$. Each item in the second batch
can increase the number of agents whose value exceeds $12$ by at
most one. Hence, at the end, at most
\[
x+19\leq 37
\]
agents have value greater than $12$, so some agent receives value
at most $12$.

The common MMS is exactly $19$. For the lower bound, partition
the items into $19$ singleton bundles containing the value-$19$
items and $19$ pairs of value-$12$ items. Every bundle has value
at least $19$. For the upper bound, there are only $57$ positive
items, each worth at most $19$. Any bundle worth strictly more
than $19$ must therefore contain at least two positive items.
A partition into $38$ such bundles would require at least $76$
positive items, which is impossible.

\medskip
\noindent
\textbf{Case 2: $x=19$.}
All $38$ items in the first batch must have been allocated as pairs
to $19$ agents. Let $E$ be the set of the other $19$ agents (the active agents), all of
whom are empty after the first batch.

The second batch consists of $38$ items of value $1$.
Let $y$ be the number of agents in $E$ whose value after this
batch is greater than $12$. Each such agent must receive at least
$13$ unit items, so
\[
y\leq \left\lfloor \frac{38}{13}\right\rfloor=2.
\]
We distinguish two subcases.

\medskip
\noindent
\textbf{Case 2(a): $y\leq 1$.}
The third batch consists of $17$ items of value $19$.

At most $17$ additional agents in $E$ can receive an item in the
third batch. Consequently, at most
\[
y+17\leq 18
\]
of the $19$ agents in $E$ can finish with value greater than $12$.
Thus some agent in $E$ receives value at most $12$.

To verify the MMS lower bound, form the following $38$ bundles:
\begin{itemize}
    \item $17$ singleton bundles containing the value-$19$ items;
    \item $19$ pairs of value-$12$ items;
    \item two bundles, each containing $19$ unit items.
\end{itemize}
Every bundle has value at least $19$, so the common MMS is at
least $19$.

We next prove that it is at most $19$. Suppose, for a contradiction,
that the items admit a partition into $38$ bundles, each worth
strictly more than $19$. Since all item values are integers, each
bundle must have value at least $20$.

We may assume that the $17$ value-$19$ items belong to distinct
bundles. Indeed, if a bundle contains at least two such items,
there is another bundle containing no value-$19$ item. The latter
bundle contains a positive item of value $s\in\{1,12\}$.
Exchange that item with one of the value-$19$ items in the former
bundle. After the exchange, the former bundle has value at least
\[
38-19+s\geq 20,
\]
while the latter bundle has value at least
\[
20-s+19\geq 20.
\]
This exchange increases the number of bundles containing a
value-$19$ item. Repeating it yields the asserted arrangement.

Let $p$ be the number of these $17$ bundles that also contain at
least one value-$12$ item. Each of the remaining $17-p$ bundles
must contain at least one unit item to reach value $20$.
Thus these bundles consume at least $17-p$ unit items.

Consider the other $21$ bundles, which contain no value-$19$ item.
For $j=1,\ldots,21$, let $a_j$ and $b_j$ denote the numbers of
value-$12$ items and unit items, respectively, in the $j$th such
bundle. At least $p$ of the $38$ value-$12$ items are already used
in bundles containing a value-$19$ item, so
\[
\sum_{j=1}^{21} a_j\leq 38-p.
\]
Moreover, the requirement $12a_j+b_j\geq 20$ implies
\[
b_j\geq 8\max\{0,2-a_j\}.
\]
Indeed, if $a_j=0$, then $b_j\geq 20\geq 16$; if $a_j=1$,
then $b_j\geq 8$; and if $a_j\geq 2$, the displayed lower bound
is zero. Therefore,
\[
\begin{aligned}
\sum_{j=1}^{21} b_j
&\geq 8\sum_{j=1}^{21}\max\{0,2-a_j\}\\
&\geq 8\left(42-\sum_{j=1}^{21}a_j\right)\\
&\geq 8(4+p).
\end{aligned}
\]
Counting also the unit items in bundles containing a value-$19$
item, the partition would require at least
\[
(17-p)+8(4+p)=49+7p>38
\]
unit items. This contradicts the fact that only $38$ unit items
exist. Hence the common MMS is exactly $19$.

\medskip
\noindent
\textbf{Case 2(b): $y=2$.}
The third batch consists of $38$ items of value $6$.

Let $E'$ be the set of the $17$ agents in $E$ whose value after
the second batch is at most $12$. The other two agents in $E$
have each received at least $13$ unit items. Hence, if $t_i$
denotes the number of unit items received by agent $i\in E'$,
then
\[
\sum_{i\in E'}t_i\leq 38-2\cdot 13=12.
\]

Suppose every agent in $E'$ finishes with value greater than $12$,
and let $r_i$ be the number of value-$6$ items received by agent
$i\in E'$ in the third batch. We claim that
\[
r_i\geq 3-t_i.
\]
For $t_i=0$, at least three value-$6$ items are necessary.
For $t_i=1$, at least two are necessary.
For $t_i=2$, at least two are necessary, which is stronger than
the claimed lower bound of one.
For $t_i\geq 3$, the claimed lower bound is nonpositive and hence
holds automatically. Summing gives
\[
\sum_{i\in E'}r_i
\geq 3|E'|-\sum_{i\in E'}t_i
\geq 3\cdot 17-12
=39.
\]
This is impossible because the third batch contains only $38$
items. Therefore, some agent in $E'$ finishes with value at most
$12$.

The common MMS is exactly $19$: the items can be partitioned
into $38$ triples, each consisting of one item of each value
$12$, $1$, and $6$. Each triple is worth $19$, and the total value
of all items is $38\cdot 19$, which also gives the matching upper
bound on the MMS.

This completes the construction for $38$ agents.

\paragraph{Extension to $n>38$ agents.}
Add $n-38$ items of value $19$ to the first batch, and let $U$
be their recipients. If $|U|<n-38$, reveal $19$ additional
value-$19$ items in the second batch and a zero-valued item
in the third. At most
\[
|U|+19+19\leq n-1
\]
agents can finish above $12$: any such agent must receive
a value-$19$ item or at least two value-$12$ items.
Otherwise, $|U|=n-38$. Simulate the $38$-agent construction
on $R=N\setminus U$, virtually assigning every core item
allocated to $U$ to a fixed agent in $R$, while preserving
all allocations to $R$. Choose subsequent batches according
to these virtual allocations. Each agent in $R$ has actual
value at most its virtual value, so the construction leaves
some agent with actual value at most $12$.

Every terminal instance consists of a terminal $38$-agent
instance with MMS $19$, together with $n-38$ additional
value-$19$ items; in the first branch, the core instance is
the one from Case~1. Adding one value-$19$ item and one agent
preserves MMS $19$: extend an MMS partition by a singleton
bundle for the lower bound; conversely, from any partition
of the enlarged instance into bundles worth more than $19$,
remove the bundle containing the new item and redistribute
its remaining items among the other bundles, contradicting
the original MMS. Repeating this argument proves that every
terminal $n$-agent instance has MMS exactly $19$.
\end{proof}

\begin{lemma}\label{lem:imp-4batches}
For identical valuations and every $n\geq 3$, no online algorithm can guarantee more than a $3/5$-approximation to the MMS, even when the MMS is known to be $1$ and the items arrive in four batches.
\end{lemma}

\begin{proof}We adapt the three-agent lower-bound construction of \citet[Theorem~2.3]{online-covering97}. Although the original construction is presented for sequential arrivals, its decision tree can be implemented using at most four batches, as follows. We first describe the construction for three agents.
    
    The first batch contains two items of value $3/5$. If the algorithm assigns them to different agents, the second batch contains two items of value $1$. Regardless of their allocation, at least one agent receives total value at most $3/5$.
    
    Suppose instead that both first-batch items are assigned to the same agent. The second batch contains one item of value $1/5$ and one item of value $2/5$. If these items are assigned to different agents, the third batch contains one item of value $1$ and one item of value $1/5$. The agent who received the two first-batch items already has value $6/5$. Of the other two agents, at most one can receive the unit-valued item, while the other receives total value at most $2/5+1/5=3/5$.
    
    It remains to consider the case in which the two second-batch items are assigned to the same agent, say agent $i$. The third batch then contains one item of value $3/5$. If this item is also assigned to agent $i$, the fourth batch contains items of values $1/5$ and $2/5$. Before the fourth batch, positive value is concentrated in at most two agents, and the total value arriving in the fourth batch is only $3/5$. Hence, some agent finishes with value at most $3/5$.
    
    Otherwise, the third-batch item is assigned to an agent different from $i$. The fourth batch then contains a single item of value $1$. Before this batch, agent $i$ has received value exactly $3/5$ from the second batch, and the recipient of the third-batch item has received value $3/5$ from that batch, unless either of them is also the recipient of the first-batch items. A single additional item can raise the value of at most one of the agents whose current value is at most $3/5$. Consequently, some agent again finishes with value at most $3/5$.
    
    The MMS is exactly $1$ in every terminal instance. The four possible terminal multisets admit, respectively, the following partitions into bundles of value at least $1$: $\{1\},\{1\},\{3/5,3/5\}$;

$\{1\},\{3/5,2/5\},\{3/5,1/5,1/5\}$; 

$\{3/5,2/5\},\{3/5,2/5\},\{3/5,1/5,1/5\}$;

and
$\{1\},\{3/5,2/5\},\{3/5,3/5,1/5\}$. In the two middle cases the total value is $3$, so the MMS cannot exceed $1$. In the first case, a partition into three bundles of value strictly greater than $1$ would require two singleton bundles, which is impossible. In the last case, the unit-valued item would have to be paired with another positive-valued item, after which the remaining items cannot be divided into two bundles both having value strictly greater than $1$.
    
    For $n>3$, add $n-3$ unit-valued items to the first batch. If these items are assigned to fewer than $n-3$ distinct agents, then at most $n-4$ agents are covered by them. In each terminal multiset of the three-agent construction, at most three pairwise disjoint bundles can have value strictly greater than $3/5$. Hence some agent finishes with value at most $3/5$.

    Otherwise, let $U$ be the $n-3$ agents who receive the unit-valued items, and let $R=N\setminus U$. For the purpose of choosing later batches, whenever an item of the three-agent construction is allocated to an agent in $U$, record it virtually as allocated to an arbitrary agent in $R$; when it is allocated to an agent in $R$, record its actual recipient. The adversary follows the three-agent decision tree using this virtual allocation. Every agent in $R$ has actual value no larger than their virtual value, so the three-agent construction leaves some agent in $R$ with actual value at most $3/5$.

    Adding one unit-valued item together with one additional agent preserves an MMS of $1$. The original MMS partition can be extended by a singleton unit bundle. Conversely, if the enlarged instance admitted a partition in which every bundle had value strictly greater than $1$, the bundle containing the new unit item would also contain an old item; removing that bundle and merging its remaining old items into another bundle would yield such a partition of the original instance. Hence every terminal $n$-agent instance has MMS exactly $1$.
    
    Branches that terminate before the fourth batch are padded with batches containing only zero-valued items. Zero-valued items may also be added to equalize batch sizes, if required. Hence, the adversary uses exactly four batches and ensures that some agent receives value at most $3/5$.

\end{proof}

\begin{lemma}\label{lem:imp-6batches}
For identical valuations and every $n\geq 4$, no online algorithm can guarantee more than a $4/7$-approximation to the MMS, even when the MMS is known to be $1$ and the items arrive in six batches.
\end{lemma}

\begin{proof}We give a batched implementation of the four-agent lower bound stated by \citet[Theorem~2.4]{online-covering97}. We first describe the construction for four agents.

The first batch contains two items of value $4/7$. If the algorithm assigns them to different agents, the second batch contains three items of value $1$. Since only three agents can receive a unit-valued item, some agent finishes with value at most $4/7$.

Suppose instead that both first-batch items are assigned to the same agent. The second batch contains two items of value $2/7$. If these items are assigned to different agents, the third batch contains two items of value $1$ and two items of value $1/7$. Among the three agents other than the recipient of the first-batch items, at most two can receive a unit-valued item. Any remaining agent has received at most one item of value $2/7$ and can receive at most both items of value $1/7$. The total value is therefore at most $2/7+1/7+1/7=4/7$.

Suppose now that both second-batch items are assigned to the same agent, say agent $i$. The third batch contains one item of value $4/7$. If this item is assigned to an agent different from $i$, the fourth batch contains two items of value $1$. Before the fourth batch, at least three agents have value at most $4/7$, whereas the two unit-valued items can raise the value of at most two of them. Thus, some agent finishes with value at most $4/7$.

It remains to consider the case in which the third-batch item is also assigned to agent $i$. The fourth batch contains two items of value $2/7$. If these items are assigned to different agents, the fifth batch contains one item of value $1$ and two items of value $1/7$. Before this batch, at most two agents have value greater than $4/7$, and every other agent has received at most one fourth-batch item. The unit-valued item can cover at most one of these agents, while any other such agent can receive total value at most $2/7+1/7+1/7=4/7$.

Finally, suppose that both fourth-batch items are assigned to the same agent. The fifth batch contains one item of value $4/7$. If this item is assigned to an agent who has received no previous positive-valued item in the construction, the sixth batch contains one item of value $1$. Before the last batch, at least two agents have value at most $4/7$, and the unit-valued item can raise the value of at most one of them. Hence, some agent remains at value at most $4/7$.

Otherwise, the fifth-batch item is assigned to an agent who has already received a positive-valued item. The sixth batch contains four items of value $1/7$. Before this batch, all positive-valued items have been assigned to at most three agents. Therefore, some agent has received no item, and even if that agent receives the entire sixth batch, their final value is only $4/7$.

We next verify that the MMS is exactly $1$ in every branch. The six terminal multisets admit the following partitions into four bundles of value at least $1$:
\begin{enumerate}
\item Three items of value $1$ and two items of value $4/7$ can be partitioned as $\{1\},\{1\},\{1\},\{4/7,4/7\}$.

\item Two items of value $1$, two items of value $4/7$, two items of value $2/7$, and two items of value $1/7$ can be partitioned as $\{1\},\{1\},\{4/7,2/7,1/7\},\{4/7,2/7,1/7\}$.

\item Two items of value $1$, three items of value $4/7$, and two items of value $2/7$ can be partitioned as 

$\{1\},\{1\},\{4/7,4/7\},\{4/7,2/7,2/7\}$.

\item One item of value $1$, three items of value $4/7$, four items of value $2/7$, and two items of value $1/7$ can be partitioned as $\{1\},\{4/7,2/7,1/7\},\{4/7,2/7,1/7\},\{4/7,2/7,2/7\}$.

\item One item of value $1$, four items of value $4/7$, and four items of value $2/7$ can be partitioned as 

$\{1\},\{4/7,4/7\},\{4/7,2/7,2/7\},\{4/7,2/7,2/7\}$.

\item Four items each of values $4/7$, $2/7$, and $1/7$ can be partitioned into four bundles of the form $\{4/7,2/7,1/7\}$.

\end{enumerate}
The partitions above show that the MMS is at least $1$ in every branch. To see that it is at most $1$, note first that the total value in the second and sixth branches is $4$. In the first branch, there are only five positive-valued items, so a partition into four bundles must contain singleton bundles of value at most $1$. In the third branch, if the two unit-valued items are placed together, the remaining value is less than $3$; if they are separated, each must be paired with another item, after which the three remaining $4/7$-valued items cannot form two bundles of value greater than $1$. In the fourth branch, the unit-valued item must be paired with an item of value at least $1/7$, leaving total value at most $3$ for the other three bundles. Finally, in the fifth branch, the unit-valued item must be paired with exactly one $2/7$-valued item, since any larger addition leaves total value less than $3$. The remaining four items of value $4/7$ and three items of value $2/7$ cannot form three bundles of value greater than $1$. Thus, every terminal four-agent instance has MMS exactly $1$.

For $n>4$, add $n-4$ unit-valued items to the first batch. If they are assigned to fewer than $n-4$ distinct agents, the adversary reveals three additional unit-valued items and terminates. At most $n-5+3+1=n-1$ agents can then receive value greater than $4/7$, where the last term accounts for combining the two $4/7$-valued items. The resulting instance has MMS exactly $1$: the $n-1$ unit-valued items form singleton bundles and the two remaining items form the last bundle, while the number of positive-valued items rules out $n$ bundles of value greater than $1$.

Otherwise, let $U$ be the $n-4$ agents receiving the initial
unit-valued items, and let $R=N\setminus U$. Whenever an item of the
four-agent construction is allocated to an agent in $U$, record it
virtually as allocated to an arbitrary agent in $R$; otherwise record
its actual recipient. The adversary chooses subsequent batches according
to this virtual four-agent allocation. The actual value of every agent
in $R$ is no larger than their virtual value, so the four-agent
construction leaves some agent in $R$ with actual value at most
$4/7$.

As above, adding a unit-valued item together with one additional agent
preserves an MMS of $1$. Hence every padded terminal instance also has
MMS exactly $1$.

Branches that terminate early are padded with zero-valued items. Thus, the construction uses six batches and always leaves some agent with value at most $4/7$ while the MMS is $1$.

\end{proof}

\section{Discussion}
In this paper, we study the power of (approximate or exact) MMS advice in obtaining positive results in online settings where input arrives either in batches or item-by-item for additive valuations. We show that indeed, MMS advice enables obtaining positive results that are not obtainable without advice  or with other advice models such as max item value or total value \citep{ChenTan26,ZhouBaiWu23,inapprox-mms26}, yet more concise than value frequency advice \citep{additiniol-information25}. Our work still leaves gaps for both  heterogeneous and identical valuations, as summarized in \Cref{tab:main-results}\footnote{Some gaps for identical valuations have persisted since the 1990s \citep{online-covering97,semi-online-covering}.}.For heterogeneous valuations, our bounds determine the optimal MMS
approximation asymptotically in both arrival models. For two agents
under item-by-item arrivals, the optimal guarantee remains between
$2-\sqrt{2}$ and $12/19$. For two batches and $n\ge6$, it remains
between $4/7$ and $2/3-1/2322429$. More generally, the matching
impossibility bound for batched arrivals applies as
$\min\{k,n\}\to\infty$. Obtaining sharper bounds for smaller values
of $\min\{k,n\}$ remains open. Another natural direction would be to study other valuations such as submodular and XOS valuations, as these admit constant approximation algorithms for the offline case~\citep{FeigeGrinberg2025,AkramiEtAl2023XOS,SeddighinSeddighin2024,GhodsiEtAl2022,BarmanKrishnamurthy2020}. In the offline setting, \citet{LiSunSubadditive26} give a
$5/16$-MMS guarantee for XOS valuations and an
$\Omega(1/\log^* n)$-MMS guarantee for general subadditive
valuations.

\section*{AI Usage Statement}
Some of the proofs in this paper were developed with the assistance of OpenAI’s ChatGPT. All the results were independently verified and rewritten by the authors.

\bibliographystyle{abbrvnat}
\bibliography{references}


\appendix

\section{Two Batches without MMS Advice}
\label{app:two-batches-no-advice}

We consider nonnegative additive valuations and two batches
$B_1$ and $B_2$. All goods in $B_1$ must be allocated irrevocably
before $B_2$ is revealed. The algorithm receives no advance
information about future values, including the agents' MMS values.
The batches may have arbitrary sizes.

We show that the optimal deterministic MMS approximation ratio
in this setting is exactly $1/n$. The impossibility holds even
for identical valuations and extends to randomized algorithms
against an adaptive adversary.

\begin{proposition}
\label{prop:two-batches-no-advice}
Running round-robin in the order $1,\ldots,n$ on $B_1$ and in
the order $n,\ldots,1$ on $B_2$ guarantees
\[
v_i(A_i)\ge \frac{\mu_i}{n}
\qquad\text{for every agent }i.
\]
Conversely, no algorithm can guarantee a ratio
greater than $1/n$, even for identical valuations.
\end{proposition}

\begin{proof}
We first prove the positive guarantee.
Let $M=B_1\cup B_2$ and $m=|M|$.
Fix an agent $i$, and order all goods so that
\[
a_1\ge a_2\ge\cdots\ge a_m,
\qquad a_\ell=v_i(g_\ell).
\]
If $m<n$, then $\mu_i=0$, so the guarantee is immediate.
Otherwise, define
\[
R_i:=\sum_{k=1}^{\lfloor m/n\rfloor}a_{kn}.
\]
We show that the algorithm gives agent $i$ value at least $R_i$.

Fix $k\le\lfloor m/n\rfloor$, and let $j_k$ be the number of
goods among $g_1,\ldots,g_{kn}$ that belong to $B_1$.
The remaining $kn-j_k$ goods belong to $B_2$.

Suppose that agent $i$ has position $p$ in a round-robin order
and that the batch contains $z$ goods of value at least $a_{kn}$.
Then they receive at least
\[
\left\lfloor\frac{z+n-p}{n}\right\rfloor
\]
goods of value at least $a_{kn}$.
Indeed, before their $t$-th turn, at most $(t-1)n+p-1$ goods
have been allocated. Whenever $(t-1)n+p\le z$, at least one
such good remains, and they pick a good of at least that value.

Agent $i$ has position $i$ in $B_1$ and position $n-i+1$ in $B_2$.
Consequently, they receive at least
\[
\left\lfloor\frac{j_k+n-i}{n}\right\rfloor
+
\left\lfloor\frac{kn-j_k+i-1}{n}\right\rfloor
=k
\]
goods of value at least $a_{kn}$.
For the equality, write $j_k=cn+r$, where $0\le r<n$.
If $r<i$, the two terms are $c$ and $k-c$;
otherwise, they are $c+1$ and $k-c-1$.

Thus, the $k$-th most valuable good in $A_i$ has value at least
$a_{kn}$, and hence
\[
v_i(A_i)\ge R_i.
\]

Consider an MMS partition for agent $i$.
At least one of its $n$ bundles contains none of
$g_1,\ldots,g_{n-1}$. By nonnegativity,
\[
\mu_i\le\sum_{\ell=n}^{m}a_\ell.
\]
Partitioning this tail into consecutive blocks starting at
indices $n,2n,\ldots,\lfloor m/n\rfloor n$, each containing
at most $n$ goods, gives
\[
\sum_{\ell=n}^{m}a_\ell
\le n\sum_{k=1}^{\lfloor m/n\rfloor}a_{kn}
=nR_i.
\]
Therefore,
\[
v_i(A_i)\ge R_i\ge\frac{\mu_i}{n}.
\]

For the impossibility, assume $n\ge2$.
We adapt the classical online machine-covering construction
\citep{semi-online-covering}.
All agents have the same additive valuation.
The first batch consists of $n$ goods, each of value $1$.
After observing their allocation, the adversary chooses
the second batch as follows.

If some agent receives no good from $B_1$, let $B_2$ consist
of $n$ zero-valued goods. Every agent has MMS equal to $1$,
while the agent who received no good from $B_1$ has final
value zero.

Otherwise, every agent receives exactly one good from $B_1$.
Let $B_2$ consist of $n-1$ goods of value $n$ and one
zero-valued good. Some agent receives none of the
$n-1$ positive-valued goods in $B_2$, and therefore has
final value $1$.
The common MMS is exactly $n$: a partition can place
each value-$n$ good in a separate bundle and all $n$
unit-valued goods in the remaining bundle, while the
total value is $n^2$.

In either case, some agent receives at most $1/n$ of their MMS.
Both batches have exactly $n$ goods, so the construction
also applies when their sizes are known in advance.

\end{proof}


\section{Exact Certificate for the Two-Batch Impossibility}
\label{app:two-batch-impossibility}

The table below specifies the $24$-item three-agent gadget used in
the two-batch impossibility result. For every item $e$, the columns
$i_e,j_e,h_e$ identify its classes in the three six-part MMS
partitions; $b_e$ is its base value and $r_e,c_e,z_e$ are its
integer perturbations. The corresponding valuations are
$w_R(e)=7b_e+r_e$, $w_C(e)=7b_e+c_e$, and
$w_Z(e)=7b_e+z_e$. The reduction, the proof that these classes
form MMS partitions, and the exact exhaustive-enumeration
verification all appear in the main paper.

\begin{table}[!htbp]
\centering
\scriptsize
\setlength{\tabcolsep}{2.5pt}
\begin{tabular}{r|rrr|r|rrr}
$e$ & $i_e$ & $j_e$ & $h_e$ & $b_e$ & $r_e$ & $c_e$ & $z_e$\\
\hline
0  &0&0&0&14472& 5& 4& 3\\
1  &0&1&1& 4021&-2& 0& 1\\
2  &0&2&2& 3127&-1& 0& 0\\
3  &0&3&3&15244&-2& 1& 1\\
4  &1&2&4&18069& 0& 0& 0\\
5  &1&3&5& 1009& 0& 0& 0\\
6  &1&4&2& 1257& 0&-1& 0\\
7  &1&5&3&16529& 0& 0&-1\\
8  &2&1&3& 1477& 1& 0& 0\\
9  &2&3&1&14083& 1&-1& 0\\
10 &2&4&5&21303&-1& 0&-1\\
11 &2&5&4&    1&-1& 0& 0\\
12 &3&0&3& 3614& 0&-2& 0\\
13 &3&1&2&22038& 0& 0&-1\\
14 &3&2&5& 5240& 0& 0& 0\\
15 &3&4&0& 5972& 0& 1& 0\\
16 &4&0&4& 9466& 0&-1& 1\\
17 &4&2&1&10428& 0& 0&-1\\
18 &4&3&0& 6528& 0& 0&-2\\
19 &4&5&2&10442& 0& 0& 1\\
20 &5&0&5& 9312& 0&-1& 1\\
21 &5&1&4& 9328& 0& 0&-1\\
22 &5&4&1& 8332& 0& 0& 0\\
23 &5&5&0& 9892& 0& 0&-1
\end{tabular}
\caption{Exact integer data defining the 24-item three-agent gadget.}
\label{tab:two-batch-certificate}
\end{table}

\FloatBarrier


\end{document}